\documentclass[aps,prl,reprint,superscriptaddress]{revtex4-2}
\usepackage{booktabs}%\usepackage{caption}
\usepackage{graphicx} % for including figures
\usepackage{float}
\usepackage[utf8]{inputenc}
\usepackage[T1]{fontenc}

\usepackage{bm} % for bold math symbols
\usepackage{hyperref} % for adding hyperlinks
\usepackage{amsmath} % for advanced math typesetting
\usepackage{amssymb} % for additional mathematical symbols

\usepackage{physics}
\usepackage{xcolor}
\begin{document}

\title{Generalized Deutsch-Jozsa Algorithm for Applications in Data Classification, Logistic Regression, and Quantum Key Distribution}

\author{M. Ghadimi}
\affiliation{Deutsche Telekom Chair of Communication Networks, Technische Universität Dresden, Dresden, Germany}
\email{milad.ghadimi@tu-dresden.de}

\author{V. Salari}
\affiliation{Department of Physics and Astronomy, University of Calgary, Calgary, Alberta, Canada}
\affiliation{Institute for Quantum Science and Technology, University of Calgary, Calgary, Alberta, Canada}
%\email{doblak@ucalgary.ca}

\author{S. Bakrani}
\affiliation{Instituto de Ci\^encias Matem\'aticas e Computac\~ao (ICMC), Universidade de S\~ao Paulo, S\~ao Carlos, Brazil}
%\affiliation{Faculty of Engineering and Natural Sciences, Kadir Has University, 34083 Istanbul, Turkey}
\author{M. Zomorodi}
\affiliation{Department of Computer Science, Cracow University of Technology, Krakow, Poland}

\author{N. Gohari-Kamel}
\affiliation{Department of Physics and Astronomy, University of Calgary, Calgary, Alberta, Canada}
\affiliation{Institute for Quantum Science and Technology, University of Calgary, Calgary, Alberta, Canada}

\author{S. Moradi}
\affiliation{Department of Physics and Astronomy, University of Calgary, Calgary, Alberta, Canada}
\affiliation{Institute for Quantum Science and Technology, University of Calgary, Calgary, Alberta, Canada}

\author{D. Oblak}
\affiliation{Department of Physics and Astronomy, University of Calgary, Calgary, Alberta, Canada}
\affiliation{Institute for Quantum Science and Technology, University of Calgary, Calgary, Alberta, Canada}
\email{doblak@ucalgary.ca}

\begin{abstract}
We present a generalized Deutsch–Jozsa (DJ) quantum algorithm that not only determines both the global type of an unknown Boolean function (constant or balanced) but also determines explicit output values of the function in a single oracle query. Unlike the original DJ algorithm, which identifies only whether a function is constant or balanced, our generalization retrieves actual function output values at the same time with using a Bell state as ancilla. This makes a richer function characterization with minimal queries to have practical quantum advantages, e.g. data classification, logistic regression, and quantum cryptography.

\end{abstract}

\maketitle

%%%%%%%%%%%%%%%%%%%%%%%%%%%%%%%%%%
\section{Introduction}

Quantum computation is anticipated to address certain problems of significance more efficiently than classical computation via characteristic features of quantum systems. Since its proposal in the last decades of the twentieth century \cite{feynman2018simulating}, the attractive idea has been developed into a vigorous research field linking physics with computer science through mathematics. Deutsch provided the first quantum algorithm that showed the potential for quantum computation \cite{deutsch1985quantum, PRA1}. It was later expanded upon by Deutsch and Jozsa \cite{deutsch1992rapid}, and also improved by Cleve and co-workers \cite{cleve1998quantum}, resulting in a deterministic multi-qubit generalization known as the Deutsch-Jozsa algorithm \cite{PRA2}. Deutsch's algorithm paved the way for other more practical quantum algorithms such as Simon's period finding algorithm~\cite{simon1997power}, Shor's factorizing algorithm \cite{shor1994algorithms}, and Grover's database search algorithm \cite{grover1996fast}, all of which provide an unprecedented computational power over the existing classical algorithms. For this reason, Deutsch's algorithm is important from a theoretical viewpoint to demonstrate the power of quantum computing \cite{PRA1}, and from a demonstration viewpoint to manifest a operation of quantum computer prototypes and quantum communication and cryptography~\cite{nagata2017quantum}. DJ's algorithm has been demonstrated using nuclear magnetic resonance \cite{collins2000nmr, PRA5}, superconducting qubits \cite{wu2011experimental}, single-photon linear optics \cite{takeuchi2000experimental, PRA4, PRL}, trapped ions \cite{gulde2003implementation}, and other systems. 
In previous works, such as \cite{nagata2020some, nagata2020generalization} Deutsch's algorithm was generalized to evaluate all mappings of single-variable Boolean functions, focusing primarily on theoretical advancements without practical applications or demonstration validation. In contrast, this paper extends the Deutsch-Jozsa algorithm to a two-variable oracle, enabling simultaneous determination of function type and values.
On the other side, classification is a core task in machine learning algorithms where their goal is to classify and categorize the data based on their features. One of the promising methods in machine learning is an ensemble of classifiers, which is a method where multiple quantum models (classifiers) are trained in parallel, and their outputs are combined according to a weighing scheme to produce a final prediction \cite{schuld2015introduction, wittek2014quantum}.
Here, we generalize the Deutsch-Josza (DJ) algorithm by introducing a two-variable function, specifically for the classification of four distinct states. The structure of this paper is organized as follows: Initially, we propose quantum algorithms that generalize both the Deutsch and the Deutsch-Josza algorithms. Subsequently, we investigate some of their applications in classification tasks, akin to the quantum ensembles classifier algorithm and then we go through to quantum key distribution (QKD) protocol\cite{scarani2009security} as an application for our model with a view of security protocol for quantum communication. In the following, we conduct a symbolic simulation to facilitate a comprehensive comparison between these algorithms, versus Deutsch and Deusch-Jozsa. Finally, we summarize our findings and discuss the implications of our approach. 
%%%%%%%%%%%%%%%%%%%%%%%%%%%%%%%%%%
\section{Quantum Algorithm}

Here, we propose a quantum algorithm acting like binary classification algorithms in classical and quantum machine learning \cite{bishop2006pattern, biamonte2017quantum} with a modification of DJ Algorithm. The original Deutsch algorithm \cite{deutsch1985quantum} determines whether a boolean function is constant or balanced. Consider a Boolean function $f$ which maps a one-bit input to a one-bit output: $f:\{0,1\}\rightarrow\{0,1\}$. Deutsch algorithm determines whether $f$ is ``constant,'' meaning $f(0)=f(1)=0$ or $f(0)=f(1)=1$, or ``balanced,'' meaning $f(0)=0$ and $f(1)=1$; or $f(0)=1$ and $f(1)=0$). The original Deutsch-Josza algorithm is a single variable quantum algorithm that determines whether an $N$-input Boolean function is constant (producing the same output for all $2^N$ inputs) or balanced (producing an equal number of 0s and 1s). It achieves this in a single query, with exponential speedup over deterministic classical algorithms, which would require up to $2^{N-1}+1$ queries to guarantee the result. Here, we generalize the oracle of this algorithm for two variables, which can aid in specifying our results to the value of the function. The description is as follows:
%Here, we generalize Deutsch algorithm to determine the value of function $f$ as well. 
\subsection*{Related work and positioning}
Generalizations of the Deutsch--Jozsa (DJ) algorithm have largely followed two orthogonal axes. 
First, promise-class extensions retain DJ's “type-only” decision flavor while broadening what 
“balanced” means, for example, the weight-gap family $DJ_{n,k}$ with exact $(k{+}1)$-query optimality \cite{PRA2} where $n$ is the order of queries in $\{0,1\}^n$, and related formulations that view DJ as a special case of exact single-query tasks 
\cite{chen2020characterization}. A second strand reinterprets DJ as a building block for applications (e.g., QKD) 
or explores conceptual variations adjacent to DJ (e.g., generalizing the one-bit Deutsch case) 
\cite{nagata2010can, nagata2015deutsch, nagata2020generalization}. These lines are invaluable baselines, but they typically output only a global property bit (constant vs.\ balanced or variants thereof).
Our contribution targets a different axis: we introduce a two-register oracle that lets the circuit recover both the global DJ-type (constant/balanced) and explicit output values in essentially the same 
interference routine, trading modest resources (two data registers with a Bell ancilla) for higher information per query. This places our method as complementary to gap-promise generalizations and structural one-query characterizations rather than in direct competition. Table~\ref{tab:gdj-comparison} summarizes the distinctions in task, information returned, query complexity, oracle/register model, and typical use-cases.

\begin{table*}[t]
\squeezetable % APS helper to reduce table spacing
\caption{\textbf{Generalized Deutsch--Jozsa landscape: prior works vs.\ this paper (GDJ).}}
\label{tab:gdj-comparison}
\footnotesize
\begin{ruledtabular}
%\begin{tabular}{L{0.12\textwidth} L{0.22\textwidth} L{0.20\textwidth} L{0.12\textwidth} L{0.16\textwidth} L{0.16\textwidth}}
\begin{tabular}{p{1.2cm} p{2.2cm} p{2.0cm} p{1.2cm} p{1.6cm} p{1.6cm}}
\textbf{Reference} &
\textbf{Task} &
\textbf{Information returned (per run)} &
\textbf{Query complexity} &
\textbf{Oracle \& Register} &
\textbf{Claims} \\
\hline
\textbf{Qiu \& Zheng (2018)}\cite{PRA2} &
GDJ as a weight-gap problem (\(DJ_{n,k}\)) distinguishing constant vs.\ “\(k\)-away from balanced”. &
Type-only decision (constant vs.\ \(k\)-balanced-like). &
\(k{+}1\) queries (proved optimal). &
Standard DJ phase oracle over \(\{0,1\}^n\). &
Benchmark for gap-type generalizations. \\
\textbf{Chen, Ye \& Li (2020)}\cite{chen2020characterization} &
Characterization of all exact 1-query algorithms (incl.\ DJ) via unitary discrimination. &
When a single exact query suffices (structural). &
Not a new promise; 1-query boundary. &
Abstract black-box; DJ-style phase oracles. &
Unifying lens for compressibility to one query. \\
\textbf{Nagata \& Nakamura (2010, 2015, 2020)}\cite{nagata2017quantum, nagata2020some, nagata2020generalization} &
2010: measurement angle; 2015: DJ-inspired QKD; 2020: generalization of \emph{Deutsch}'s 1-bit case. &
QKD property bit; conceptual extensions. &
1 query for DJ decision. &
Standard DJ; crypto/entangled-state variants. &
DJ-based QKD primitive; interpretational insights. \\
\textbf{This paper (GDJ)} &
Two-register \((x,y)\) oracle; recover type and explicit values (four cases), extendable to \(2n\). &
Global type \& which of \(\{00,01,10,11\}\). &
Effective \(O(2)\) queries (vs.\ classical \(2^{2n-2}{+}1\)). &
Dual data registers + Bell ancilla; phase-kickback. &
Higher info/query; ensembles; higher-throughput QKD. \\

\end{tabular}
\end{ruledtabular}
\end{table*}

\subsection{Generalized Deutsch Algorithm}
\begin{figure*}[htbp]
\centering
\includegraphics[width=0.8\linewidth, trim={80pt 5pt 20pt 50pt}, clip]{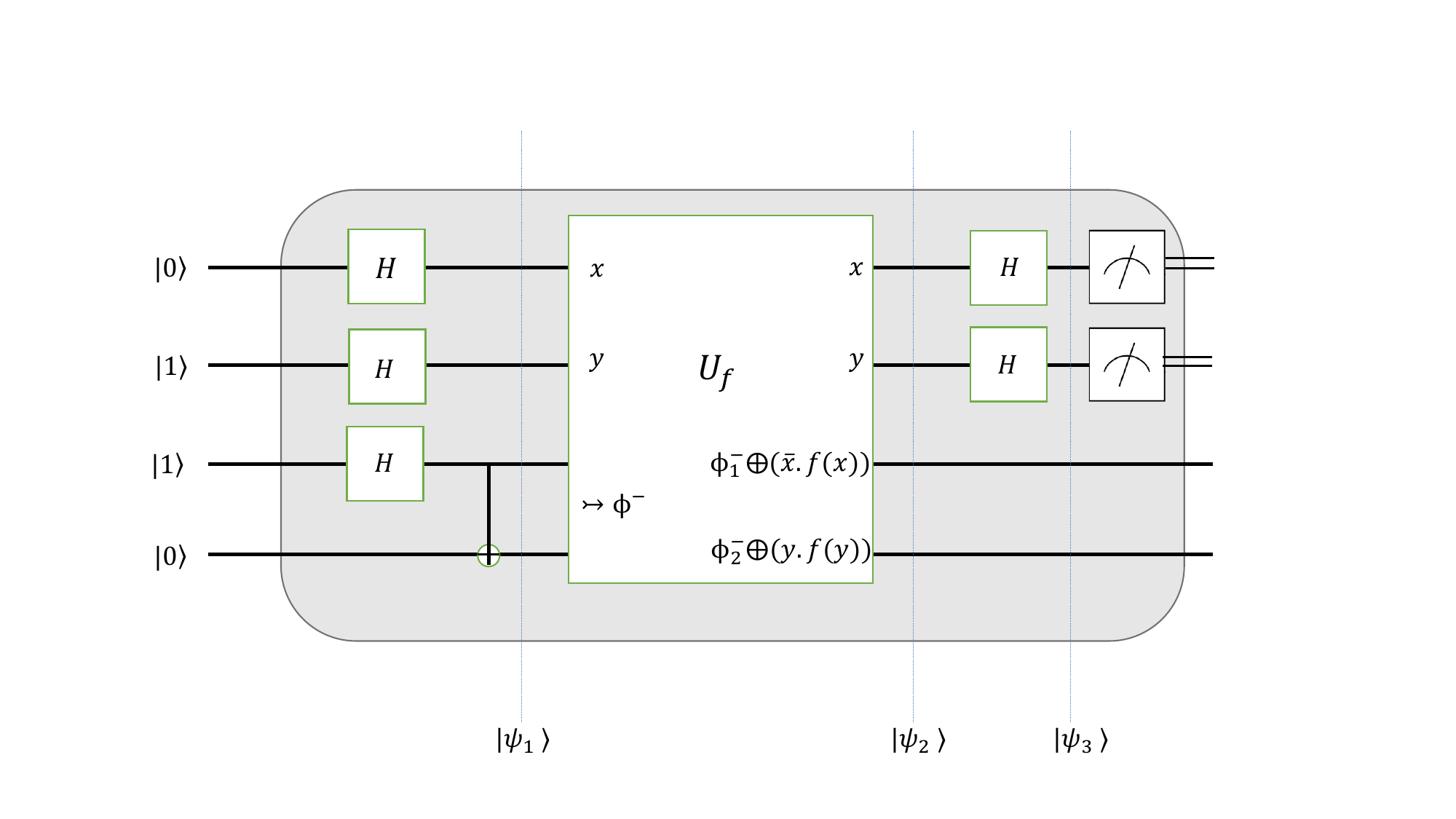}
\caption{Quantum circuit for the Generalized Deutsch Algorithm (GDA), where $\Phi^-$ denotes a Bell state, and $\Phi^-_1$ and $\Phi^-_2$ refer to the first and second qubits (or parts) of the Bell state $\Phi^-$, respectively.}\label{figc}
\end{figure*}

Our proposed Generalized Deutsch Algorithm is sketched in Fig. \ref{figc}. We are given an oracle for the quantum implementation of the function $f$ that maps: 
\begin{align}
U_{f}:~|x,y\rangle &\rightarrow(-1)^{\overline{\rm x}.f(x)}|x\rangle\ (-1)^{y.f(y)}|y\rangle \nonumber \\
&\rightarrow(-1)^{\overline{\rm x}.f(x)+y.f(y)}|x,y\rangle \label{eq1}
\end{align}
where \( x,y\in\{0,1\} \) and $\bar{x}$ is the bit-wise complement of $x$, i.e. $\bar{x}=1-x$.

To extend this formulation and leverage quantum entanglement for more efficient computation, we introduce an ancilla register in the Bell state \( |\Phi^-\rangle \). The Bell states are a set of four maximally entangled two-qubit states that form an orthonormal basis for the two-qubit Hilbert space. Specifically, the state \( |\Phi^-\rangle \) is defined as
\[
|\Phi^-\rangle = \frac{1}{\sqrt{2}} \left( |00\rangle - |11\rangle \right),
\]
Using this ancilla allows us to encode the function's evaluation into phase differences via entanglement, which is particularly useful in quantum algorithms because the orthonormality of the Bell states preserves distinguishability and enables interference effects to capture the function’s behavior in a more information-rich (and potentially parallelized) manner compared to classical bits.
\\
The application of \(U_{f}\) on \(|x,y\rangle|\Phi^{-}\rangle\) yields
\begin{align}
U_{f}:~|x,y\rangle|\Phi^{-}\rangle &\rightarrow |x,y\rangle|\Phi^{-} \oplus (f(x,y)), \Phi^{-}\oplus (f(x,y))\rangle \nonumber \\
&\rightarrow |x, y\rangle |\Phi_{1}^{-} \oplus (\overline{x} \cdot f(x)), \Phi_{2}^{-} \oplus (y \cdot f( y))\rangle \label{eq2}
\end{align} 
where the subscripts 1 and 2 denote the first and second qubits of the state \( |\Phi^-\rangle \), and \(\oplus\) and \(.\) denote addition modulo two and multiplication, respectively. Here, \( |\Phi^{-}\rangle \) represents ancilla qubits used to store partial results. Since the four Bell states are orthonormal, the use of the Bell-state ancilla lets us capture the function’s effect in a more information-rich way.
 Here, we employ a completed version of the circuit for simulating a quantum circuit on a quantum computer. These tools are crucial for developing and advancing quantum computing, allowing researchers and developers to study the potential of quantum algorithms and understand their implications for various applications, from cryptography to complex system modeling.
Fig. \ref{figc} shows the quantum circuit, and as demonstrated, there are two main qubits and a $|-\rangle$ ancillary. The initial state of the algorithm-after passing the Hadamard gates-is denoted as $|\psi_{1}\rangle$. Subsequently, applying the oracle to $|\psi_{1}\rangle$ yields $|\psi_{2}\rangle$, where the oracle's computation is executed on ancillary qubits. Upon simplifying $|\psi_{2}\rangle$, we obtain Eq.~\ref{eq.4}. Passing it through two additional Hadamard gates, we arrive at $|\psi_{3}\rangle$ Eq.~\ref{eq.5}, representing the system's final state. The steps are as follows: 
\begin{equation}
|\psi_{1}\rangle=\frac{1}{\sqrt{2^3}}\big(|0\rangle+|1\rangle)\big(|0\rangle-|1\rangle)\big(|00\rangle-|11\rangle)
\end{equation}
The Oracle operates on the state as:
\begin{multline}
|\psi_{2}\rangle = \frac{1}{\sqrt{2^3}} \Big( |00\rangle|0 \oplus (\overline{0} \cdot f(0)), 0 \oplus ({0} \cdot f(0))\rangle \\
- |10\rangle|0 \oplus (\overline{1} \cdot f(1)), 0 \oplus ({0} \cdot f(0))\rangle \\
+ |01\rangle\vert 0 \oplus (\overline{0} \cdot f(0)), 0 \oplus ({1} \cdot f(1))\rangle \\
- |11\rangle|0 \oplus (\overline{1} \cdot f(1)), 0 \oplus ({1} \cdot f(1))\rangle \Big) \\
- \Big( |00\rangle|1 \oplus (\overline{0} \cdot f(0)), 1 \oplus ({0} \cdot f(0))\rangle \\
- |10\rangle|1 \oplus (\overline{1} \cdot f(1)), 1 \oplus ({0} \cdot f(0))\rangle \\
+ |01\rangle|1 \oplus (\overline{0} \cdot f(0)), 1 \oplus ({1} \cdot f(1))\rangle \\
- |11\rangle|1 \oplus (\overline{1} \cdot f(1)), 1 \oplus ({1} \cdot f(1))\rangle \Big)\label{eq.4}
\end{multline}

After the oracle, the state can be written as follows by extracting the kets:
\begin{multline}
|\psi_{2}\rangle = \frac{1}{\sqrt{2^3}} \Big( (-1)^{f(0)}|00\rangle|00\rangle - |10\rangle|00\rangle \\
+ (-1)^{f(0)+f(1)}|01\rangle\vert 00\rangle - (-1)^{f(1)}|11\rangle|00\rangle \Big) \\
- \Big( (-1)^{f(0)}|00\rangle|11\rangle - |10\rangle|11\rangle \\
+ (-1)^{f(0)+f(1)}|01\rangle|11\rangle - (-1)^{f(1)}|11\rangle|11\rangle \Big)
\end{multline}

Then the final state after Hadamard operations (see Fig.~\ref{figc}) is:
%\begin{widetext}
\begin{equation}
\begin{split}
&\vert \psi_{3}\rangle=\frac{1}{\sqrt{2^5}}\left[\left((-1)^{f(0)}-1\right)\vert 1\rangle+\left(1+(-1)^{f(0)}\right)\vert 0\rangle\right]\\
&\otimes\left[\left(1-(-1)^{f(1)}\right)\vert 0\rangle+\left(1+(-1)^{f(1)}\right)\vert 1\rangle\right] (|00\rangle-|11\rangle)\label{eq.5}
\end{split}
\end{equation}
%\end{widetext}

Therefore, by performing measurements on the register, one can obtain one of the possible outcomes as follows: If we get $|0\rangle$ from the first line and $|1\rangle$ from the second line, then $f(0)=f(1)=0$ and $f$ is constant as $f=0$. If we get $|0\rangle$ from the first line and $|0\rangle$ from the second line, then $f(0)=0$ and $f(1)=1$, thus $f$ is balanced. If we get $|1\rangle$ from the first line and $|1\rangle$ from the second line, then $f(0)=1$ and $f(1)=0$, thus $f$ is balanced. If we get $|1\rangle$ from the first line and $|0\rangle$ from the second line, then $f(0)=f(1)=1$ and $f$ is constant as $f=1$. The benefit of the GDA over DA is that it can discriminate the constant (balanced) functions in four different states, while DA can only discriminate two states (constant or balanced). In other words, DA cannot discriminate two states ($f(0)=f(1)=1$) and ($f(0)=f(1)=0$) for constant function, and similarly, it cannot discriminate states ($f(0)=0$ and $f(1)=1$) and ($f(0)=1$ and $f(1)=0$) for balanced function.

\subsection{Generalized Deutsch-Jozsa (DJ) algorithm}
\begin{figure*}[htbp]
\centering
\includegraphics[width=0.8\linewidth, trim={80pt 5pt 20pt 50pt}, clip]{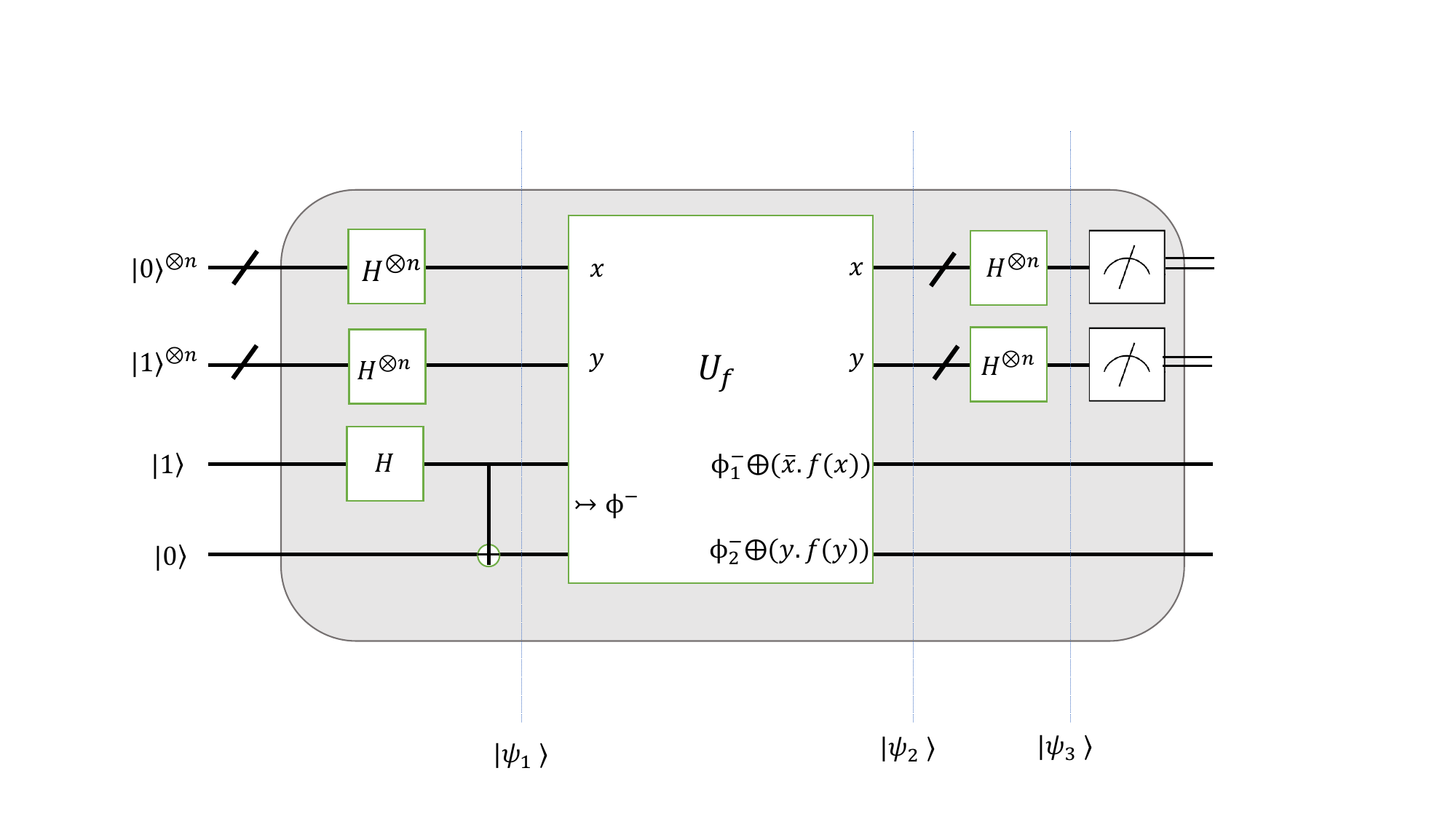}
\caption{Quantum Circuit of Generalized Deutsch-Jozsa Algorithm (GDJA).}\label{figc2}
\vspace{-10pt}
\end{figure*}

The Deutsch-Jozsa (DJ) algorithm is exponentially faster than any possible deterministic classical algorithm, in which it takes n-digit binary values as input and single digits 0 or 1 as output. Similar to the Deutsch algorithm, the function is either constant (0 on all inputs or one on all inputs) or balanced (returns 1 for half of the input domain and 0 for the other half); the task then is to determine if the function $f
$ is constant or balanced. Here, we would like to generalize the DJ algorithm as well, according to the circuit sketched in FIG.\ref{figc2}, to obtain the value of the function besides the nature of the function (balanced or constant) to classify the inputs.\\
We suppose that the inputs are $2n$-digit registers in which the initial state of the system is $|0\rangle^{\otimes n} |1\rangle ^{\otimes n}$.
We are given the same oracle for quantum implementation of the function $f$ that maps according to equation 1, but for registers $x$ and $y$ each one has length n, i.e. $x,y\in\{0,1\}^{2n}$. Again, in this protocol, we can determine the value of the function besides the type of the function (see appendix for the running the algorithm on an IBM quantum computer).

The quantum circuit of algorithm is depicted in Fig.~\ref{figc2}. In this case, the input of the system is $|0\rangle^{\otimes n} |1\rangle ^{\otimes n}$, where after passing through Hadamard gates $H^{\otimes 2n}$, it becomes:
\begin{equation}
|\psi_1\rangle=\frac{1}{\sqrt{2^2}}\big(|0\rangle+|1\rangle\big)^{\otimes n}\big(|0\rangle-|1\rangle\big)^{\otimes n}\big(|00\rangle-|11\rangle\big)
\end{equation}
which can be written in the form of Fourier transforms as:
\begin{equation}
\sum_{x \in \lbrace {0,1}\rbrace^n}^{N} (-1)^{i\cdot x} |x\rangle \sum_{y \in \lbrace {0,1}\rbrace^n} ^{N}  (-1)^{j\cdot y} |y\rangle \big(\frac{|00\rangle-|11\rangle}{\sqrt{2}}\big) 
\end{equation}
For example, when the inputs are $|0\rangle |1\rangle$, it means $i=0$ and $j=1$, thus we have:
\begin{equation}
\sum_{x \in \lbrace {0,1}\rbrace^n}^{N} |x\rangle \sum_{y \in \lbrace {0,1}\rbrace^n} ^{N}  (-1)^{y} |y\rangle \big(\frac{|00\rangle-|11\rangle}{\sqrt{2}}\big)  
\end{equation}

The oracle function $f$ acts as equation 1 for $x,y\in\{0,1\}^{2n}$. The state after the oracle is:
\begin{equation}
|\psi_2\rangle=\sum_{x \in \lbrace {0,1}\rbrace^n}^{N} (-1)^{\overline{\rm x}\cdot f(x)} |x\rangle \sum_{y \in \lbrace {0,1}\rbrace^n} ^{N}   (-1)^{y+y\cdot f(y)} |y\rangle  
\end{equation}

After the application of Hadamard gates $H^{\otimes 2n}$:
\begin{equation}
\begin{split}
   &|\psi_3\rangle=\sum_{x \in \lbrace {0,1}\rbrace^n}^{N} (-1)^{\overline{\rm x}\cdot f(x)} \sum (-1)^{x\cdot i} |i\rangle \\
   &\times \sum_{y \in \lbrace {0,1}\rbrace^n} ^{N}   (-1)^{y+y\cdot f(y)}\sum (-1)^{y\cdot j} |j\rangle (\frac{|00\rangle-|11\rangle}{\sqrt{2}}) 
\end{split}
\end{equation}

We should obtain $i$ and $j$ with probabilities $p_i=|\phi(i)|^2$ and $p_j=|\varphi(j)|^2$, where $\phi(i)=\sum_{x \in \{0,1\}^n}^{N} (-1)^{\overline{\rm x}\cdot f(x)+x\cdot i}$ and $\varphi(j)=\sum_{y \in \lbrace {0,1}\rbrace^n} ^{N} (-1)^{y+y\cdot f(y)+y\cdot j}$. Finally, by measuring the computational basis on the register, we have the results below:

For $i=0$ and $j=1$, we have a general result for the algorithm as follows:

\begin{equation}
\phi(0)=\sum_{x \in \{0,1\}^n}^{N} (-1)^{\overline{\rm x}\cdot f(x)} 
\end{equation}
and
\begin{equation}
\varphi(1)=\sum_{y \in \lbrace {0,1}\rbrace^n} ^{N} (-1)^{y+y\cdot f(y)+y}
\end{equation}

If $f(x)=0$ and $f(y)=0$, we obtain:
\begin{equation}
\phi(0)=\sum_{x \in \lbrace {0,1}\rbrace^n}^{N} 1 \quad \text{and} \quad \varphi(1)=\sum_{y \in \lbrace {0,1}\rbrace^n} ^{N}  1
\end{equation}
which means $f$ is constant. If $f(x)=1$ and $f(y)=0$, we obtain:
\begin{equation}
\phi(0)=0 \quad \text{and} \quad \varphi(1)=\sum_{y \in \lbrace {0,1}\rbrace^n} ^{N}  1
\end{equation}
therefore, $f$ is balanced. If $f(x)=0$ and $f(y)=1$, we obtain:
\begin{equation}
\phi(0)=\sum_{x \in \lbrace {0,1}\rbrace^n}^{N} 1 \quad \text{and} \quad \varphi(1)=\sum_{y \in \lbrace {0,1}\rbrace^n} ^{N}   (-1)^{3y} =0
\end{equation}
where it means $f$ is balanced.

If $f(x)=1$ and $f(y)=1$, we obtain:
\begin{equation}
\phi(0)=\sum_{x \in \lbrace {0,1}\rbrace^n}^{N} (-1)^{\overline{\rm x}}=0 \quad \text{and} \quad \varphi(1)=\sum_{y \in \lbrace {0,1}\rbrace^n} ^{N}   (-1)^{3y}=0
\end{equation}
which means $f$ is constant.

We always obtain $i= 0$ and $j=1$ when $f(x)=0$ and $f(y)=1$, but we never obtain $i= 0$ and $j=1$ when $f(x)=1$ and $f(y)=0$.

\begin{figure*} 
   \centering
    \includegraphics[scale=0.60]{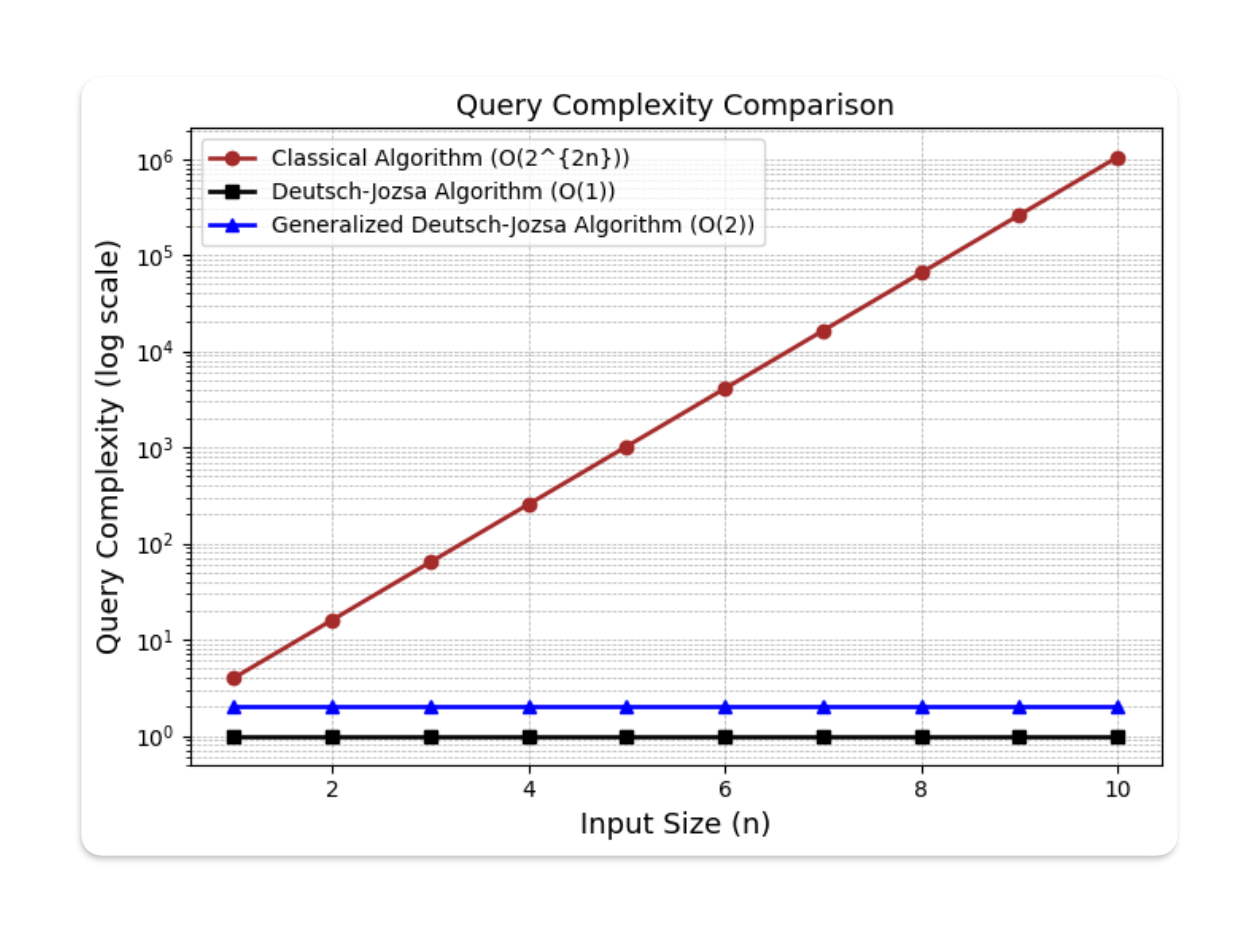}
    \caption{Query complexity for classical and quantum algorithms in computations based on determining a constant or balanced function.}
    \label{fig:query-complexity}
\end{figure*}

%\subsection{Query complexity}
Query complexity refers to the number of calls the algorithm makes to an oracle (or black box function) \cite{aaronson2021open, chen2020characterization}.
In the classic scenario, if we consider using this type of oracle for GDJ algorithm, we would require $2^{n-2}+1$ queries. However, in this context, the problem can be solved with certainty using only 2 queries (see Table1 and Fig.~\ref{fig:query-complexity}). %(with the fourth state being predictable).

\begin{table}[htbp]
\centering
\caption{A comparison of the domain and query size of algorithms.}
\label{tab:deutsch}
\begin{tabular}{|c|c|c|}
\hline
\textbf{ALGORITHM} & \textbf{Domain} & \textbf{Query} \\
\hline
Deutsch & $\{0,1\} \rightarrow \{0,1\}$ & $\mathcal{O}(1)$ \\
\hline
Generalized Deutsch & $\{0,1\}^2 \rightarrow \{0,1\}^2$ & $\mathcal{O}(2)$ \\
\hline
Deutsch-Jozsa & $\{0,1\}^n \rightarrow \{0,1\}$ & $\mathcal{O}(1)$ \\
\hline
Generalized Deutsch-Jozsa & $\{0,1\}^{2n} \rightarrow \{0,1\}^2$ & $\mathcal{O}(2)$ \\
\hline
\end{tabular}
\end{table}

% \iffalse
% \begin{table}[htbp]
% \label{table1}
% \begin{center}
% \begin{tabular}{ lcc|cc} 
% \caption{A comparison of the domain and query size of algorithms.}
% \hline
% ALGORITHM & Domain \\
% \hline
% \hline
% DA & $ 
% \lbrace  {0,1} \rbrace \rightarrow \lbrace {0,1}\rbrace $ \\
% \hline
% GDA & $\lbrace {0,1}\rbrace^2 
% \rightarrow \lbrace {0,1}\rbrace^2 $ \\
% \hline
% DJA & $\lbrace {0,1}\rbrace^n \rightarrow\lbrace {0,1}\rbrace $ \\
% \hline
% GDJA & $\lbrace {0,1}\rbrace^{2n} \rightarrow \lbrace {0,1}\rbrace^2 $ \\
% \hline
% \end{tabular}\label{table2}
% \end{center}
% \end{table}
% \fi
% \begin{table}
% \begin{center}
% \begin{tabular}{ |c|c|c|c| } 
% \hline
% ALGORITHM & Domain & Query\\
% \hline
% \hline
%  Deutsch & $ 
% \lbrace  {0,1} \rbrace \rightarrow \lbrace {0,1}\rbrace $ & $O(1)$ \\
% \hline
% Generalized Deutsch & $\lbrace {0,1}\rbrace^2 
% \rightarrow \lbrace {0,1}\rbrace^2 $ & $O(2)$ \\
% \hline
% Deutsch-Jozsa & $\lbrace {0,1}\rbrace^n \rightarrow\lbrace {0,1}\rbrace $ & $O(1)$ \\
% \hline

% Generalized Deutsch-Jozsa & $\lbrace {0,1}\rbrace^{2n} \rightarrow \lbrace {0,1}\rbrace^2 $ & $O(2)$ \\
% \hline
% \end{tabular}\label{table2}
% \end{center}
% \end{table}

%%%%%%%%%%%%%%%%%%%%%%%%%%%%%%%%%%%%%%%%

%%%%%%%%%%%%%%%%%%%%%%%%%%%%%%%%%%%%%%%%%

\subsubsection{Life-Death Rooms as a Simulation for Oracles} 
\begin{figure*}[htbp]
% If you need the images to span two columns in a two-column document, use figure*:
%\begin{figure*}[ht!]

\includegraphics[scale=0.4]{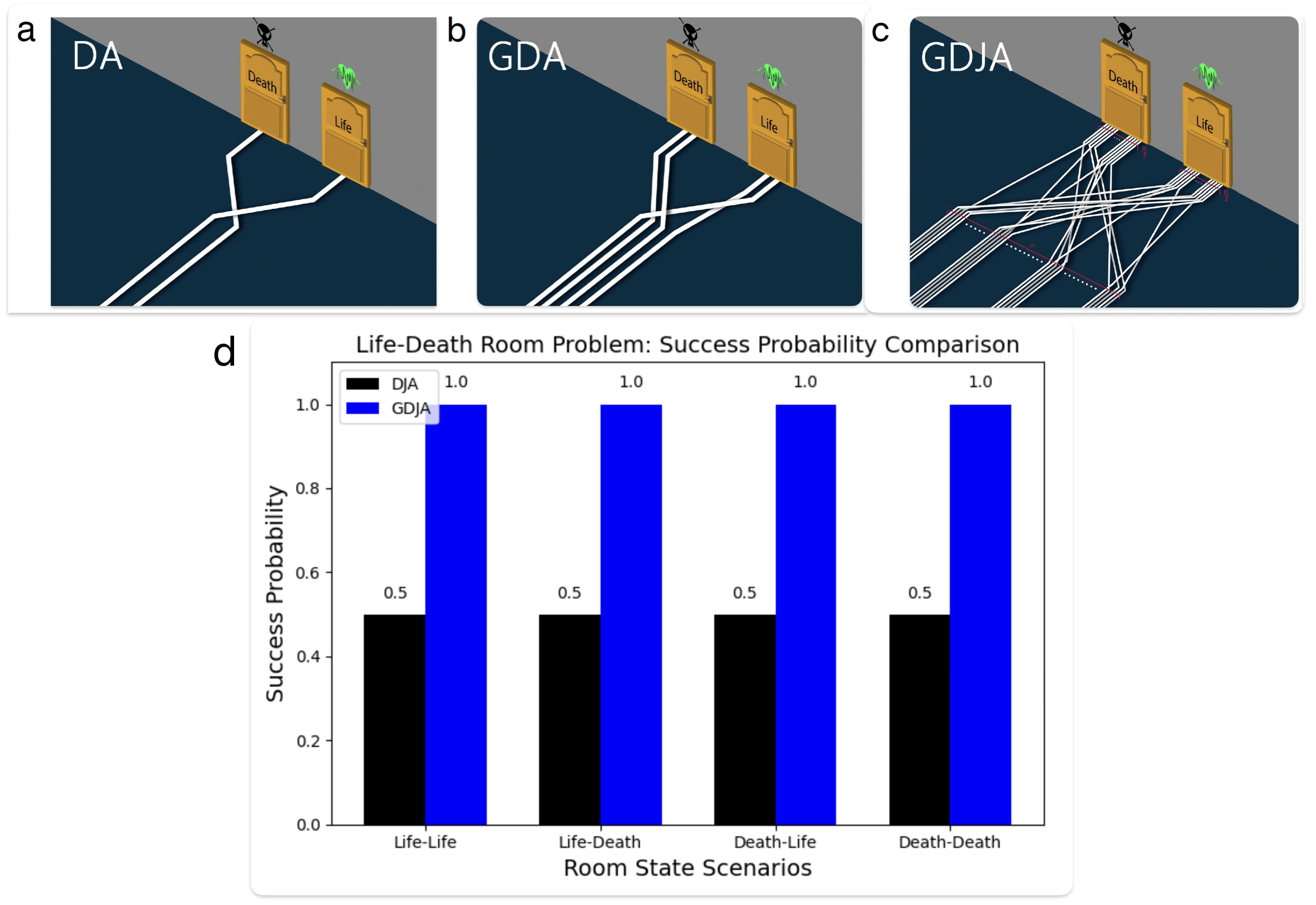}
\caption{Simulation of the oracles used in (a) the Deutsch algorithm (DA), (b) the generalized Deutsch algorithm (GDA), and (c) the generalized Deutsch–Jozsa algorithm (GDJA) through a “Life–Death choice game.” In this analogy, different input pathways lead to two possible outcome rooms: “Life” or “Death”, representing the binary outputs of the oracle. The objective of the game is to identify which paths lead to which rooms, i.e., to uncover the true nature of the output states. (d) Four possible configurations exist for the two outcome rooms: (1) Life–Death, (2) Life–Life, (3) Death–Life, and (4) Death–Death. When viewed as a game, the generalized algorithms guarantee a winning probability of 1, since they can fully determine both the nature of the doors (outputs) and the mapping of the input pathways.}
\label{game}
%\end{figure*} % Use this if you started with figure*
\end{figure*}

Here, we would like to make a symbolic simulation to show the oracle (i.e. GDJA) relative to DA and DJA. In fact, here we simulate the oracles as an issue of determination of the nature of two rooms that we don't know what is going on in different pathways (e.g. in a maze) that approach to only two rooms, each room with only two possible states of "death" and "life" as depicted in Fig.~\ref{game}. We only know that there are four possibilities for the rooms: Death-Death (e.g. $\lbrace  {0,0} \rbrace$), Death-Life (e.g. $\lbrace  {0,1} \rbrace$), Life-Death (e.g. $\lbrace  {1,0} \rbrace$), and Life-Life (e.g. $\lbrace  {1,1} \rbrace$) as the values of the function in the oracle. In fact, the normal Deutsch and DJ's algorithms only determine that whether the rooms are similar or different but they don't specify which one is death-room and which one is the life-room. However, the generalized algorithms, GDA and GDJA, can determine which door is life or death. For example, based on GDJA, it can be used as a game in which there are $n$ ways to the two doors and we intend to know which ways goes to the life rooms to survive, otherwise the game is over. However, regarding this simulation as a game, the probability of winning (i.e., specification of life rooms) via GDJ algorithm (GDJA) is 1 as the nature of doors and the ways going to the rooms can be determined (see Fig.~\ref{game}d).

\section{Practical Advantages}

\subsection{Quantum Classifiers}

In machine learning, an ensemble method combines multiple models to enhance overall predictive performance. This approach can be extended to quantum computing through the creation of quantum classifier ensembles. Quantum ensembles leverage properties of quantum states, such as superposition and entanglement, to enable parallel evaluation of multiple classifier models~\cite{abbas2020quantum}. These models are typically weighted based on their accuracy or other performance metrics.

\begin{figure*}
    \centering
    \includegraphics[scale=0.59]{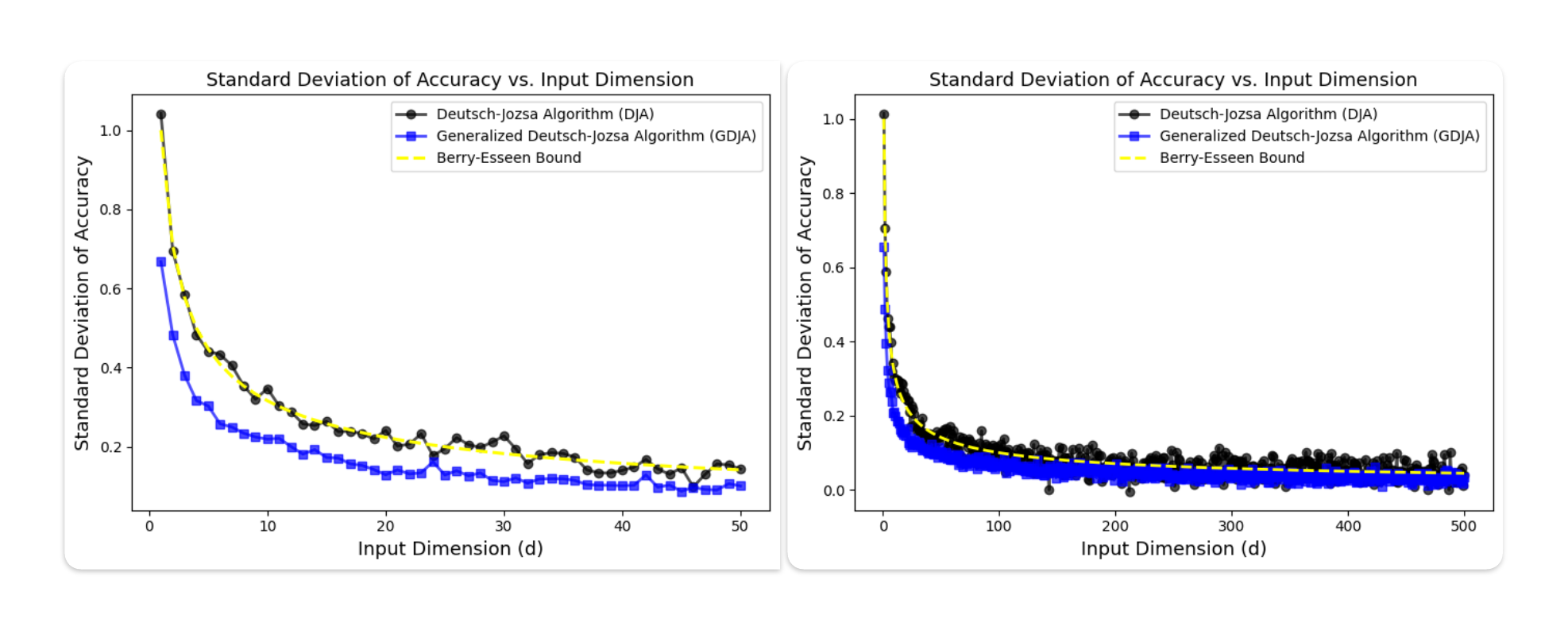}
    \caption{Standard deviation of accuracy versus dimension for the DJ and GDJ algorithms, shown for dimensions up to 50 (left) and 500 (right). The red dashed line represents the Berry-Esseen bound as a theoretical reference for convergence.}
    \label{fig:EST}
\end{figure*}

In the context of quantum ensembles, we adapt the Generalized Deutsch-Jozsa (GDJ) framework to evaluate and weight classifiers. Here, each classifier's parameters are encoded into quantum states $\ket{\theta_i, \phi_i}$, and its prediction on an input $(x, y)$ is represented as a binary function $f_{\theta, \phi}(x, y)$. The GDJ framework emphasizes the encoding of features and parameters, which is central to the classification process. Below, we outline the steps for constructing an ensemble classifier using this algorithm, focusing on encoding parameters and inputs.

\subsubsection{Feature Encoding}
In the GDJ framework, features $(x, y)$, where $x, y \in \{0, 1\}$ represent binary attributes of a data sample, are encoded as quantum states $\ket{x, y}$. These serve as direct inputs to the oracle $U_f$, which evaluates the function $f(x, y)$. The state $\ket{x, y}$ captures the data in superposition, allowing parallel computation across all possible inputs.

For general $d$-dimensional binary features, we use $n = \lceil \log_2 d \rceil$ qubits to represent a feature vector $\mathbf{x} = (x_1, \ldots, x_d)$ as $\ket{\mathbf{x}}$ and $\mathbf{y} = (y_1, \ldots, y_d)$ as $\ket{\mathbf{y}}$, prepared in superposition:
\begin{equation}
\ket{\psi_{\text{feat}}} = \frac{1}{\sqrt{2^n}} \sum_{\mathbf{x,y}} \ket{\mathbf{x,y}}.
\end{equation}

This enables parallel evaluation over all inputs via the oracle $U_f$.

\subsubsection{Parameter Encoding}
The parameters $(\theta, \phi)$, such as thresholds or weights, define the behavior of the classifier functions $f(x)$ and $f(y)$. These are embedded within the oracle's logic rather than directly encoded into the quantum state $\ket{x, y}$. They influence the evaluation of $f(x, y)$ by determining phase shifts during the oracle's transformation.

The GDJ oracle operates as equation 1, where $f(x)$ and $f(y)$ are binary outputs governed by $\theta, \phi$. For a binary classifier, $f(\mathbf{x})$ might be defined as $f(\mathbf{x}) = \Theta(\theta \cdot \mathbf{x} - \phi)$, where $\Theta$ is the Heaviside step function.

\subsubsection{Combining Features and Parameters}
The oracle $U_f$ integrates features and parameters by applying phase shifts based on $f(x)$ and $f(y)$. For the full operation, including an ancillary qubit $\ket{\Phi^{-}} = \frac{1}{\sqrt{2}} (\ket{00} - \ket{11})$ for phase kickback:

\begin{equation}
U_f: \ket{\mathbf{x,y}} \ket{\Phi^{-}} \rightarrow (-1)^{\overline{x} \cdot f(\mathbf{x})+y\cdot f(\mathbf{y})} \ket{\mathbf{x,y}} \ket{\Phi^{-}}.
\end{equation}

This encodes the classifier's decision into the phase, leveraging interference to distinguish function types (constant or balanced) and values in a single query.

\subsubsection{Ensemble Encoding}
To evaluate $N$ classifiers in parallel, their parameters $(\theta_i, \phi_i)$ are encoded in superposition:

\begin{equation}
\ket{\Psi_{\text{param}}} = \frac{1}{\sqrt{N}} \sum_{i=1}^N \ket{\theta_i, \phi_i},
\end{equation}

assuming discretizations (e.g., using $m$ qubits per parameter). The combined state is:

\begin{equation}
\ket{\Psi} = \ket{\psi_{\text{feat}}} \otimes \ket{\Psi_{\text{param}}} = \frac{1}{\sqrt{2^n N}} \sum_{\mathbf{x,y}, i} \ket{\mathbf{x,y}} \ket{\theta_i, \phi_i}.
\end{equation}

\subsubsection{Classifier Evaluation}
The oracle evaluates $f_{\theta_i, \phi_i}(\mathbf{x,y})$:
\begin{align}
\begin{split}
U_f \ket{\mathbf{x,y}} \ket{\theta_i, \phi_i} \ket{\Phi^{-}} \rightarrow \\(-1)^{\overline{x} \cdot f_{\theta_i, \phi_i}(\mathbf{x})+y\cdot f_{\theta_i, \phi_i}(\mathbf{y})} \ket{\mathbf{x,y}} \ket{\theta_i, \phi_i} \ket{\Phi^{-}}
\end{split}
\end{align}
Phase kickback encodes the output in the phase for interference-based classification.

\subsubsection{Weighting Scheme}
Classifiers are weighted by accuracy $w_{\theta_i, \phi_i} \in [0, 1]$ via controlled rotations:

\begin{equation}
\ket{\theta_i, \phi_i} \rightarrow \sqrt{w_{\theta_i, \phi_i}} \ket{\theta_i, \phi_i} + \sqrt{1 - w_{\theta_i, \phi_i}} \ket{\text{aux}},
\end{equation}

where $\ket{\text{aux}}$ is discarded. After normalization (scaling weights so $\sum_i w_{\theta_i, \phi_i} = 1$) and Hadamard gates on the feature register, the state is:

\begin{equation}
\ket{\Psi'} = \frac{1}{\sqrt{N}} \sum_{\mathbf{x}, i} \sqrt{w_{\theta_i, \phi_i}} \ket{\mathbf{x}} \ket{\theta_i, \phi_i} \ket{f_{\theta_i, \phi_i}(\mathbf{x})}.
\end{equation}

\subsubsection{Measurement and Probabilities}
The probability for class $c \in C = \{00, 10, 01, 11\}$ (for the two-variable case) is:

\begin{equation}
p(f(\mathbf{x}) = c) = \sum_{\theta_i, \phi_i \in E_c} w_{\theta_i, \phi_i},
\end{equation}

where $E_c = \{(\theta_i, \phi_i) \mid f_{\theta_i, \phi_i}(\mathbf{x}) = c\}$.

\subsubsection{Decision Making}
The final class is:
\begin{equation}
c^* = \arg\max_{c \in C} p(f(\mathbf{x}) = c),
\end{equation}

with random selection in case of ties.

%\subsubsection{Comparison with Quantum Ensemble Methods}
%Like other quantum ensembles that use superposition for parallel evaluation, the GDJ framework employs interference for efficient processing. However, the GDJ oracle uniquely determines both function type and value in one query, enhancing interpretability and efficiency. This repurposes the GDJ mechanism for weighted decision-making in ensembles, potentially solving classically hard problems via quantum speedup.

%Figure~\ref{fig:EST} compares the standard deviation of accuracy for the original Deutsch-Jozsa (DJ) and GDJ algorithms. The DJ oracle exhibits significant fluctuations due to its simplicity, while the GDJ oracle, with its dual registers, shows higher initial variability but smoother, faster convergence toward the Berry-Esseen bound \cite{Berry1941, Esseen1942, Petrov1995}. This indicates how added complexity and registers improve stability in higher dimensions.

\begin{figure*}
    \centering
    \includegraphics[scale=0.6]{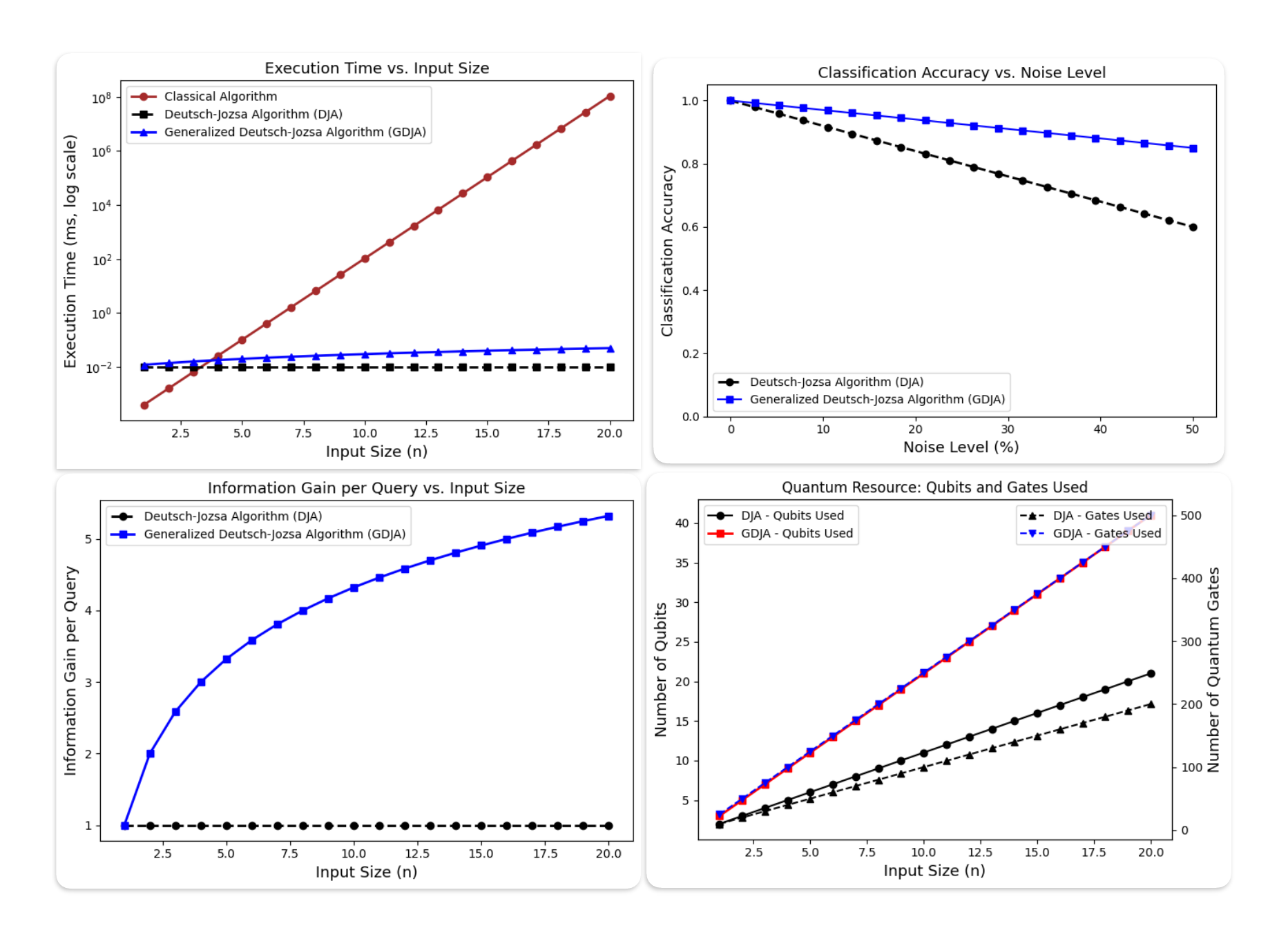}
    \caption{Top-left: Execution time versus input size for classical, DJ, and GDJ algorithms. Top-right: Classification accuracy versus noise level for DJ and GDJ algorithms. Bottom-left: Information gain per query versus input size for DJ and GDJ algorithms. Bottom-right: Quantum resource usage (qubits and gates) versus input size for DJ and GDJ algorithms.}
    \label{fig:execution}
\end{figure*}

\subsubsection{Comparison with Quantum Ensemble Methods}

Like other quantum ensembles that use superposition for parallel evaluation, the GDJ framework employs interference for efficient processing. However, the GDJ oracle uniquely determines both function type and value in one query, enhancing interpretability and efficiency. This repurposes the GDJ mechanism for weighted decision-making in ensembles, potentially solving classically hard problems via quantum speedup.

Figure~\ref{fig:EST} compares the standard deviation of accuracy for the original DJ and GDJ algorithms. The DJ oracle exhibits significant fluctuations due to its simplicity, whereas the GDJ oracle, with its dual registers, shows higher initial variability but smoother, faster convergence toward the Berry–Esseen bound \cite{Berry1941,Esseen1942}. Recall that the Berry–Esseen theorem states that, for a sum of \(n\) independent and identically distributed (i.i.d.)\ random variables \(X_i\) with mean \(0\), variance \(\sigma^2\), and third absolute moment \(\rho = \mathbb{E}[|X_1|^3]\), the cumulative distribution function \(F_n(x)\) of the normalized sum satisfies
\begin{equation}
\sup_x \left| F_n(x) - \Phi(x) \right| \le \frac{C\,\rho}{\sigma^3\sqrt{n}},
\end{equation}
where \(C\) is a universal constant (\(C \approx 0.4748\) in the best known bound~\cite{Petrov1995}). In our context, the bound provides an asymptotic limit on the rate at which the outcome distribution of the quantum algorithm approaches the Gaussian distribution. The smoother convergence of GDJ toward this limit indicates that added register complexity and oracle structure enhance statistical stability in higher dimensions.

\subsection{Quantum Machine Learning and Ensemble Methods}

%\begin{figure*}
    %\centering
   %\includegraphics[scale=0.5]{Fig.QKD.pdf}
    %\caption{Key generation rate vs input dimension for DJA and GDJA for a potential application in QKD.}
    %\label{figqkd}
%\end{figure*}

In classical machine learning, ensembles aggregate predictions from multiple classifiers for improved robustness. Quantum computers enable more efficient implementations by encoding classifiers in superposition and evaluating them in parallel.

\subsubsection{Quantum Ensemble via GDJ Oracle}
For $B$ classifiers, prepare the superposition:
\begin{equation}
\frac{1}{\sqrt{B}} \sum_{j=1}^{B} \ket{j} \; \ket{x, y} \; \ket{0},
\end{equation}

where $\ket{j}$ indexes the classifier, $\ket{x, y}$ encodes the input, and the third register holds the output.

Apply operations in parallel:
\begin{equation}
\frac{1}{\sqrt{B}} \sum_{j=1}^{B} \ket{j} \; \ket{x, y} \; \ket{f_j(x, y)}.
\end{equation}

The GDJ oracle combines results via interference, e.g., to compute majority votes.

\subsubsection{Efficiency Improvement}
Quantum construction requires $O(\log B)$ operations for superposition, versus $O(B)$ classically. Circuit depth matches a single classifier plus $\mathrm{poly}(\log B)$ overhead.

\subsubsection{A Toy Example}
Consider three quantum classifiers (A, B, C), each voting on whether a binary input $(x, y) = (0, 1)$ is classified as positive (1) or negative (0). For simplicity, we map the pair $(f(x), f(y))$ to a single class, e.g., positive if $f(x) \vee f(y) = 1$. Their individual predictions are:
\[
\begin{aligned}
f_A(0) &= 1, \quad f_A(1) = 0 \quad \text{(positive)}, \\
f_B(0) &= 0, \quad f_B(1) = 0 \quad \text{(negative)}, \\
f_C(0) &= 1, \quad f_C(1) = 1 \quad \text{(positive)}.
\end{aligned}
\]
The ensemble quantum state after encoding these results is:
\begin{equation}
\frac{1}{\sqrt{3}}\Big( \; \ket{A}\ket{1} + \ket{B}\ket{0} + \ket{C} \ket{1} \; \Big).
\end{equation}

By applying our GDJ-based interference oracle, the amplitude of the outcome corresponding to ``1'' is amplified. Upon measurement, the state is higher for ``positive,'' reflecting the majority vote in a single quantum evaluation.

\subsubsection{Noise Sensitivity in DJA vs. GDJA}
The DJA applies a single oracle query followed by an interference-based measurement, making it highly sensitive to noise. Even small errors in quantum gates or measurements can significantly impact classification accuracy \cite{Preskill2018}. A bit-flip error causes incorrect measurement collapse \cite{NielsenChuang}, leading to a linear accuracy degradation modeled as $A_{DJA}(\eta) = 1 - \beta \eta$, where $\eta$ is the noise level and $\beta$ reflects DJA’s high noise sensitivity. Since DJA relies on a single oracle call and final measurement, each noise event directly affects the outcome. The GDJA improves robustness by using multiple registers and an extended oracle, enabling partial error correction and noise distribution across qubits. This results in a quadratic accuracy degradation $A_{GDJA}(\eta) = 1 - \gamma \eta^2$,  where $\gamma < \beta$. Errors accumulate gradually rather than collapsing the outcome, making GDJA significantly more resilient in practical quantum computing environments \cite{Bharti2022} (see Fig.~\ref{fig:execution}).

\subsubsection{Information Gain per Query vs. Input Size}

The efficiency of quantum algorithms can be evaluated by the information gain per query, particularly in decision problems like DJA and GDJA. GDJA provides a more detailed classification, leading to higher information gain per query. Information gain per query, $I(n)$, is the number of useful bits extracted per oracle query. DJA performs binary classification (constant vs. balanced), yielding a fixed information gain $I_{DJA}(n) = 1 \text{ bit per query.}$ GDJA, in contrast, retrieves additional function values, resulting in a logarithmic increase $I_{GDJA}(n) = 1 + \log_2(n) \text{ bits per query.}$ Since DJA is limited to binary classification, its information gain remains constant, whereas GDJA scales logarithmically, enhancing its utility for larger inputs. The growing information gain in GDJA enables more complex classifications and decision-making within a single query, reinforcing its advantages in quantum information processing (see Fig.~\ref{fig:execution}).

\subsubsection{Quantum Resources: Qubits and Gates Used}

The efficiency of a quantum algorithm is often evaluated based on the number of qubits and quantum gates required for its implementation \cite{NielsenChuang}. The DJA and GDJA differ significantly in their resource requirements due to their distinct computational structures. In DJA, the number of qubits required is given by $Q_{DJA}(n) = n + 1$ where $n$ is the number of input bits. The additional qubit is used as an ancillary qubit for phase kickback. For GDJA, since it extends DJA by processing more function information, it requires additional input registers $Q_{GDJA}(n) = 2n + 1$.
This additional register allows GDJA to retrieve more than just the binary classification of functions, thus increasing its qubit demand. The number of quantum gates in an algorithm determines its depth and computational cost. DJA employs a relatively small number of quantum gates, given by $G_{DJA}(n) = O(n)$ which includes Hadamard gates for superposition, an oracle query, and additional measurement operations. For GDJA, the complexity increases due to the need for an extended oracle and additional function evaluation steps $G_{GDJA}(n) = O(n^2)$.

%The increased qubit and gate count in GDJA reflect its capability to handle more complex function evaluations. While DJA remains highly efficient in terms of resource usage, its utility is limited to binary function classification. GDJA, on the other hand, demands more resources but provides more valuable outputs, making it better suited for advanced quantum decision problems.

DJA requires fewer qubits and gates, making it lightweight and efficient but limited in its output. GDJA demands more qubits and gates, scaling as $O(n^2)$ in gate complexity but providing richer classification capabilities. The trade-off between efficiency and information extraction suggests that GDJA is advantageous when additional function details are needed (see Fig.~\ref{fig:execution}).

\subsection{As a Cryptography Protocol}

The GDJ algorithm can be directly adapted as a black-box primitive for quantum key distribution (QKD) between two parties Alice and Bob.  
Let Alice choose a two-variable Boolean function 
\[
f: \{0,1\}^n \times \{0,1\}^n \to \{0,1\}
\]
of one of four possible types:  
\[
\text{constant-0},\ \text{constant-1},\ \text{balanced-01},\ \text{balanced-10}.
\]
She keeps \(f\) secret from an eavesdropper (Eve). The protocol can be written in the following steps:
\begin{enumerate}
    \item \textit{State preparation:} Alice prepares the \(2n+1\)-qubit state
    \[
    |\psi_{\text{in}}\rangle = |0\rangle^{\otimes n} |1\rangle^{\otimes n} \otimes |\Phi^{-}\rangle,
    \]
    %where 
    %\(|\Phi^{-}\rangle = \frac{1}{\sqrt{2}} (|00\rangle - |11\rangle)\) 
    %is a Bell state used for phase kickback.
    \item \textit{Oracle application:} Alice sends the state through the GDJ oracle $U_f$ which encodes the function type (constant or balanced \cite{nagata2015deutsch, nagata2010can}) and its explicit values in a single query.
    \item \textit{Transmission to Bob:} The output state
    \[
    |\psi_{\text{out}}\rangle = U_f |\psi_{\text{in}}\rangle
    \]
    is transmitted to Bob over a quantum channel.
    \item \textit{Measurement:} Bob applies Hadamard gates \(H^{\otimes 2n}\) to the first two registers and measures in the computational basis to obtain \((i,j)\). The mapping between outcomes \((i,j)\) and function types is deterministic where $(i,j) \in \{(0,1), (0,0), (1,1), (1,0)\}$ giving specific $f$ type and values.
    
\end{enumerate}

\begin{figure*}
    \centering
    \includegraphics[scale=0.5]{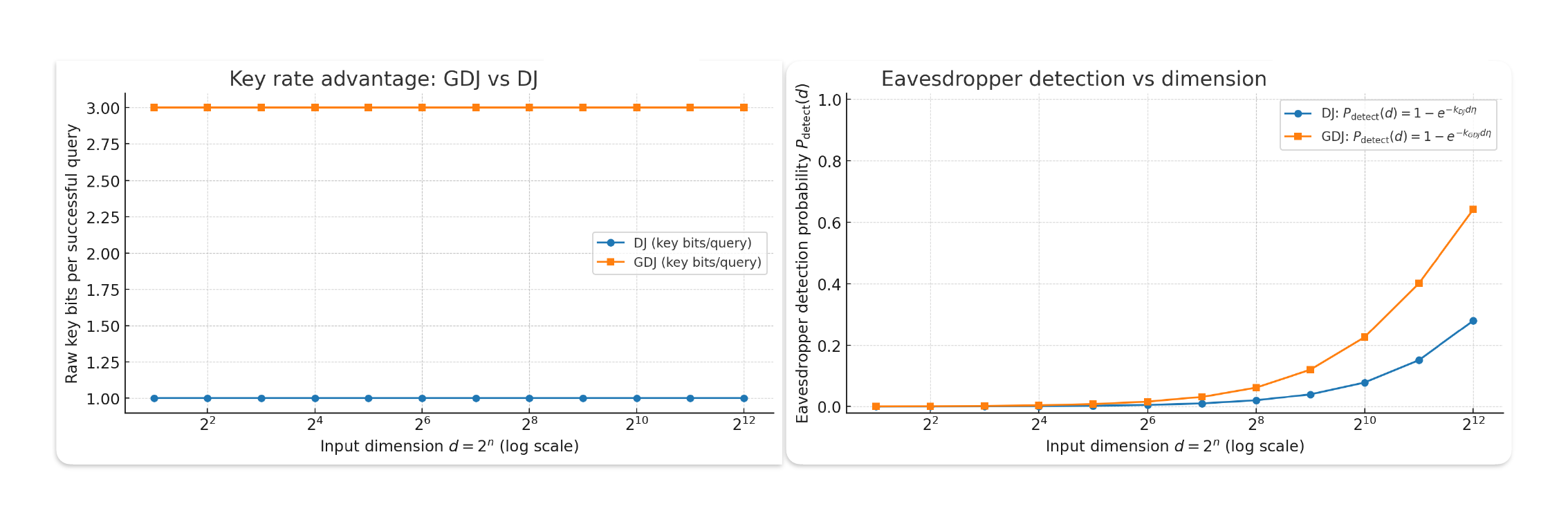}
    \caption{Comparison of GDJ and DJ algorithms for QKD. 
Left: Raw key rate per successful query vs.\ input dimension \(d=2^n\), showing GDJ’s constant threefold improvement (\(R_{\mathrm{GDJ}}=3\) vs.\ \(R_{\mathrm{DJ}}=1\)). 
Right: Eavesdropper detection probability \(P_{\mathrm{detect}}(d)=1-e^{-k\,d\,\eta}\) with \(\eta=0.1\), \(k_{\mathrm{DJ}}=8\times 10^{-4}\), and \(k_{\mathrm{GDJ}}=2.5\times 10^{-3}\), illustrating GDJ’s faster rise with \(d\).}
    \label{fig:GDJ_QKD_advantages}
\end{figure*}

Assuming Alice and Bob are running QKD session across $d$ transmissions, on each transmission, Bob’s checker either \emph{raises a flag} (an outcome inconsistent with the honest interference pattern) or it does not. We model this per-trial outcome by a Bernoulli variable $Z_t\in\{0,1\}$ with $Z_t\sim\mathrm{Bern}(q)$ \cite{Kawakami2017Bernoulli}. A Bernoulli random variable $Z$ takes values in $\{0,1\}$ with $\Pr(Z=1)=q$ and $\Pr(Z=0)=1-q$. Its \textit{mean} and \textit{variance} are $\mathbb{E}[Z]=q$ and $\mathrm{Var}(Z)=q(1-q)$, respectively \cite{Kawakami2017Bernoulli, Maeda2019FiniteKey, Koashi2020Tutorial}. Our per-trial outcome is binary (flag/no-flag), so $Z_t\sim\mathrm{Bern}(q)$ is the natural model for each transmission.
%$\mathbb{E}[Z_t]$ is the \emph{expectation} (mean) of the per-trial flag variable $Z_t\in\{0,1\}$. For a Bernoulli($q$) variable, }

%\paragraph{\textcolor{red}{$\mathbb{E}[Z_t]$}}
%\textcolor{red}{ is the \emph{expectation} (mean) of the per-trial flag variable $Z_t\in\{0,1\}$. For a Bernoulli($q$) variable,}
%\[
%\textcolor{red}{\mathbb{E}[Z_t] \;=\; 0\cdot (1-q) + 1\cdot q \;=\; q \;=\; \Pr(Z_t=1).}
%\]
So, for an indicator $Z_t=\mathbf{1}\{\text{flag on trial }t\}$ we have $\mathbb{E}[Z_t]=\Pr(\text{flag on trial }t)$. In our model:
\[
\mathbb{E}[Z_t]=q_0 \text{ (honest channel)}, \mathbb{E}[Z_t]=q_1 \text{ (attacked trial)}
\] where $q_0$ and $q_1$ are flag rates, in which $q_0$ = per-trial flag probability with no eavesdropper; and $q_1>q_0$ is under attack (information–disturbance). The gap $q_1-q_0$ quantifies disturbance and drives the detection exponent \cite{FuchsPeres1996,Scarani2009RMP}.
Over $d$ trials, the total number of flags $S=\sum_{t=1}^d Z_t$ is a simple count, binomial under independent and identically distributed $Z_t$\ assumptions \cite{CoverThomas2006}, which lets Bob apply standard hypothesis tests, and the sample mean $\bar Z=\frac{1}{d}\sum_{t=1}^d Z_t$ is an unbiased estimator of the flag rate ($\mathbb{E}[\bar Z]=q$) with variance $q(1-q)/d$. In the honest channel $q=q_0$; if Eve performs intercept–resend on that trial, $q=q_1>q_0$ (information–disturbance). Eve measures and re-sends, erasing phase relations. The baseline attack that makes post-Hadamard outcomes nearly uniform over the admissible symbols, letting us compute per-trial catch $k$, e.g., $1/2$ for DJ, $3/4$ for GDJ (see Appendix). Given the probability a single attacked trial is flagged is $k$, and $\eta$ is the fraction of the $d$ transmissions that Eve tampers with ($m=\eta d$). Under independence over $m$ attacked trials, for small $k$ and large $m$, the flag count is approximated by $\mathrm{Poisson}(k m)$, where $(1-k)^m\approx e^{-k m}$, giving $P_{\mathrm{detect}}(d)\approx 1-e^{-k\,\eta d}$\cite{CoverThomas2006}. So, $\eta$ scales the number of informative trials and thus the exponent in the detection law. "i.i.d" means independence across trials and the same law per trial. It makes binomial/Poisson counting models and sharp large-deviation (Chernoff/Hoeffding) bounds for the miss probability (see Appendix).

%\noindent\textcolor{red}{\textbf{Poisson approximation.} For small $\alpha$ and large $m$, the flag count $S\sim\mathrm{Bin}(m,\alpha)$ is approximated by $\mathrm{Poisson}(\alpha m)$. It yields $P_{\mathrm{detect}}(d)=1-e^{-\alpha\,\eta d}$ in one line, matching Eq.~(31).}

\paragraph{Security intuition.}
Any attempt by Eve to measure or clone the transmitted state will, by the no-cloning theorem \cite{wootters1982single}, introduce disturbance into the interference pattern at Bob's end.  
Let \(d = 2^n\) as the input dimension \cite{bennett1984bb84}. So, if Eve measures a fraction \(\eta\) of the qubits, the detection probability is
\begin{equation}
P_{\mathrm{detect}}(d) = 1 - e^{-k d \eta},
\end{equation}
where \(k > 0\) depends on the physical implementation (channel loss, detector efficiency, etc.).  
This exponential form reflects that as \(d\) increases, Eve’s disturbance becomes exponentially easier to detect, i.e.,
\[
\lim_{d \to \infty} P_{\mathrm{detect}}(d) = 1.
\]

\paragraph{Key generation rate.}
In each successful run, Alice and Bob learn one bit from the function type and an additional \(\log_2 4 = 2\) bits from its explicit value:
\[
R_{\mathrm{key}} = 1 + 2 = 3 \quad \text{bits per successful query}.
\]
Compared to the standard DJ-based QKD protocol (yielding only 1 bit per query), the GDJ-based protocol increases the raw key rate by a factor of \(3\) while preserving unconditional security against intercept–resend attacks.

Figure~\ref{fig:GDJ_QKD_advantages} illustrates two key advantages of the GDJ algorithm over the standard DJ algorithm when applied to QKD. In the left panel, the raw key rate (in bits per successful query) is shown as a function of the input dimension \(d = 2^n\) for \(n \in \{1,\dots,12\}\) on a logarithmic scale. For DJ, each query yields exactly one bit (\(R_{\mathrm{DJ}} = 1\)), corresponding solely to the binary classification of the function as constant or balanced. In contrast, GDJ deterministically extracts both the function type and its explicit output values, yielding \(R_{\mathrm{GDJ}} = 3\) bits per query, a constant threefold improvement in the raw key generation rate. The right panel compares the eavesdropper detection probability, modeled as Eq.31. 
In our illustrative example, we set $\eta = 0.1$, meaning that Eve disturbs $10\%$ of the transmitted quantum signals. This represents a moderate ``attack strength'' often used to visualize detection curves, neither extreme nor trivial, and is reasonable for demonstration purposes. The parameters $k_{\mathrm{DJ}}$ and $k_{\mathrm{GDJ}}$ are phenomenological sensitivity constants that encapsulate the combined effects of hardware visibility, gate and measurement noise, and the degree to which the protocol’s interference pattern amplifies disturbance. Varying these constants shifts the detection curves horizontally, but the qualitative conclusion, that GDJ rises faster than DJ, remains unchanged. With the chosen parameters, the detection probability in Eq.31
crosses key waypoints at physically sensible dimensions. For a $50\%$ detection probability, solving $e^{-k\eta d} = 0.5$ gives
$d_{50} = \frac{\ln 2}{k \eta}$. For DJ, $k\eta = 8 \times 10^{-5}$ yields $d_{50} \approx 8{,}663$ ($n \approx 13.1$), while for GDJ, $k\eta = 2.5 \times 10^{-4}$ yields $d_{50} \approx 2{,}772$ ($n \approx 11.4$). Similarly, for a $90\%$ detection probability, $d_{90} = \frac{\ln 10}{k \eta}$, we find $d_{90} \approx 28{,}783$ ($n \approx 14.8$) for DJ and $d_{90} \approx 9{,}210$ ($n \approx 13.2$) for GDJ. These thresholds indicate that GDJ achieves high detection probabilities at significantly smaller input dimensions than DJ, consistent with the security advantage predicted by our model. In Fig.~\ref{fig:GDJ_QKD_advantages}, we set \(\eta = 0.1\), \(k_{\mathrm{DJ}} = 8\times 10^{-4}\), and \(k_{\mathrm{GDJ}} = 2.5\times 10^{-3}\), resulting in a significantly steeper rise for GDJ with increasing dimension \(d\). So, GDJ’s dual-register oracle structure amplifies disturbance effects, making eavesdropping exponentially easier to detect as the system dimension grows. 
\subsubsection{Quantum advantage for QKD}
The GDJ–based QKD protocol makes a distinct advantage compared to conventional QKD protocols such as BB84 \cite{bennett1984bb84}, B92\cite{bennett1992b92}, and E91\cite{ekert1991e91}. Traditional quantum schemes rely on multiple transmission bases or entangled pairs to exchange single bits per quantum state, but the GDJ algorithm deterministically retrieves both the function type (constant/balanced) and the explicit output values within a single oracle query, resulting in a threefold increase in key generation rate (3 bits/query). Its dual-register entangled oracle enhances information density and allows simultaneous encoding of phase and value information, improving efficiency without extra queries. Moreover, the GDJ structure yields an exponentially higher eavesdropper detection probability, since any intercept–resend attack disrupts the interference pattern more strongly than in traditional protocols. The GDJ algorithm is a simple gate-level structure, it relies only on Hadamard and CNOT operations and a single Bell-state ancilla, making it straightforward to implement on current quantum devices with practical feasibility on today’s NISQ hardware while still delivering a threefold key-rate improvement and exponentially faster eavesdropper detection with a high-performance yet hardware-ready framework for quantum-secure communication.\\
The security of the GDJ-based QKD protocol is dual in nature:
(i) \emph{information-theoretic}, guaranteed by the no-cloning theorem \cite{wootters1982single}, any eavesdropping introduces measurable disturbance; and
(ii) \emph{computational-theoretic}, arising from the GDJ oracle’s complexity, which encodes two-register phase correlations inaccessible to an adversary.
Consequently, Eve cannot reconstruct the interference pattern without causing detectable disturbance. Fig.~7 demonstrates an example that contrasts GDJ’s and DJ’s detection probabilities. By embedding QKD in the GDJ framework, the protocol achieves deterministic identification of both function type and value in one query, higher key rates (\(3\times\) over standard DJ-based schemes), and exponentially increasing eavesdropper detection probability with system dimension \(d\). This makes GDJ a useful protocol for high-dimensional, interference-based QKD systems.

\section{Conclusion}

In this work, we introduced the Generalized Deutsch--Jozsa (GDJ) algorithm, an extension of the original Deutsch--Jozsa framework that enables the simultaneous determination of a Boolean function’s type (constant or balanced) and its explicit output values. This dual capability significantly enhances the algorithm’s utility in practical quantum information tasks. We showed that GDJ achieves a query complexity of $O(4)$, compared to the $O(2^{2n})$ queries required classically for functions with $2n$ inputs, thus preserving the exponential speedup of the original algorithm while providing richer information per query. The additional output value retrieval makes GDJ particularly well-suited for applications such as multi-class quantum classification, ensemble-based quantum machine learning, and high-dimensional quantum key distribution, where both efficiency and interpretability are essential. Our results demonstrate that careful algorithmic modifications, such as the use of dual registers and enhanced oracle structures, can yield quantum protocols that are theoretically faster and may better aligned with the demands of real-world quantum technologies.

\vspace{0.2cm}
\noindent
\textbf{Acknowledgments} ---
MG acknowledges the German Research Foundation (DFG, Deutsche Forschungsgemeinschaft) as part of Germany’s Excellence Strategy – EXC2050/1 – Project ID 390696704 – Cluster of Excellence “Centre for Tactile Internet with Human-in-the-Loop” (CeTI) at Technische Universität Dresden. VS and DO acknowledge support from the National Research Council of Canada and the Natural Sciences and Engineering Research Council through its Discovery Grant Program, the Alliance Quantum Consortia Grant, and the New Frontiers in Research Fund in Canada. SB acknowledges support from FAPESP (grant no. 2023/04294-0). All authors acknowledge the use of IBM Quantum services for this work.
%Discussions with F.T. Ghahramani and Arash Tirandaz are greatly acknowledged. The authors acknowledge support from Spanish MINECO/FEDER FIS2015-69983-P, Basque Government IT986-16, and the projects QMiCS (820505) and OpenSuperQ (820363) of the EU Flagship on Quantum Technologies. This material is also based upon work supported by the U.S. Department of Energy, Office of Science, Office of Advance Scientific Computing Research (ASCR), under field work proposal number ERKJ335.

\vspace{0.2cm}

\bibliographystyle{apsrev4-2}
\bibliography{bib}

@incollection{feynman2018simulating,
  title={Simulating physics with computers},
  author={Feynman, Richard P},
  booktitle={Feynman and computation},
  pages={133--153},
  year={2018},
  publisher={CRC Press}
}

@article{deutsch1985quantum,
  title={Quantum theory, the Church--Turing principle and the universal quantum computer},
  author={Deutsch, David},
  journal={Proceedings of the Royal Society of London. A. Mathematical and Physical Sciences},
  volume={400},
  number={1818},
  pages={97--117},
  year={1985},
  publisher={The Royal Society London}
}

@article{deutsch1992rapid,
  title={Rapid solution of problems by quantum computation},
  author={Deutsch, David and Jozsa, Richard},
  journal={Proceedings of the Royal Society of London. Series A: Mathematical and Physical Sciences},
  volume={439},
  number={1907},
  pages={553--558},
  year={1992},
  publisher={The Royal Society London}
}

@article{cleve1998quantum,
  title={Quantum algorithms revisited},
  author={Cleve, Richard and Ekert, Artur and Macchiavello, Chiara and Mosca, Michele},
  journal={Proceedings of the Royal Society of London. Series A: Mathematical, Physical and Engineering Sciences},
  volume={454},
  number={1969},
  pages={339--354},
  year={1998},
  publisher={The Royal Society}
}

@article{simon1997power,
  title={On the power of quantum computation},
  author={Simon, Daniel R},
  journal={SIAM journal on computing},
  volume={26},
  number={5},
  pages={1474--1483},
  year={1997},
  publisher={SIAM}
}

@inproceedings{shor1994algorithms,
  title={Algorithms for quantum computation: discrete logarithms and factoring},
  author={Shor, Peter W},
  booktitle={Proceedings 35th annual symposium on foundations of computer science},
  pages={124--134},
  year={1994},
  organization={Ieee}
}

@inproceedings{grover1996fast,
  title={A fast quantum mechanical algorithm for database search},
  author={Grover, Lov K},
  booktitle={Proceedings of the twenty-eighth annual ACM symposium on Theory of computing},
  pages={212--219},
  year={1996}
}

@article{PRA1,
  title = {Deutsch-Jozsa algorithm as a test of quantum computation},
  author = {Collins, David and Kim, K. W. and Holton, W. C.},
  journal = {Phys. Rev. A},
  volume = {58},
  issue = {3},
  pages = {R1633--R1636},
  numpages = {0},
  year = {1998},
  month = {Sep},
  publisher = {American Physical Society},
  doi = {10.1103/PhysRevA.58.R1633},
  url = {https://link.aps.org/doi/10.1103/PhysRevA.58.R1633}
}

@article{PRA2,
  title = {Generalized Deutsch-Jozsa problem and the optimal quantum algorithm},
  author = {Qiu, Daowen and Zheng, Shenggen},
  journal = {Phys. Rev. A},
  volume = {97},
  issue = {6},
  pages = {062331},
  numpages = {9},
  year = {2018},
  month = {Jun},
  publisher = {American Physical Society},
  doi = {10.1103/PhysRevA.97.062331},
  url = {https://link.aps.org/doi/10.1103/PhysRevA.97.062331}
}

@article{PRA4,
  title = {Implementing Deutsch-Jozsa algorithm using light shifts and atomic ensembles},
  author = {Dasgupta, Shubhrangshu and Biswas, Asoka and Agarwal, G. S.},
  journal = {Phys. Rev. A},
  volume = {71},
  issue = {1},
  pages = {012333},
  numpages = {8},
  year = {2005},
  month = {Jan},
  publisher = {American Physical Society},
  doi = {10.1103/PhysRevA.71.012333},
  url = {https://link.aps.org/doi/10.1103/PhysRevA.71.012333}
}

@article{PRA5,
  title = {Implementation of the refined Deutsch-Jozsa algorithm on a three-bit NMR quantum computer},
  author = {Kim, Jaehyun and Lee, Jae-Seung and Lee, Soonchil and Cheong, Chaejoon},
  journal = {Phys. Rev. A},
  volume = {62},
  issue = {2},
  pages = {022312},
  numpages = {4},
  year = {2000},
  month = {Jul},
  publisher = {American Physical Society},
  doi = {10.1103/PhysRevA.62.022312},
  url = {https://link.aps.org/doi/10.1103/PhysRevA.62.022312}
}

@article{PRL,
  title = {Deutsch-Jozsa Algorithm Using Triggered Single Photons from a Single Quantum Dot},
  author = {Scholz, M. and Aichele, T. and Ramelow, S. and Benson, O.},
  journal = {Phys. Rev. Lett.},
  volume = {96},
  issue = {18},
  pages = {180501},
  numpages = {4},
  year = {2006},
  month = {May},
  publisher = {American Physical Society},
  doi = {10.1103/PhysRevLett.96.180501},
  url = {https://link.aps.org/doi/10.1103/PhysRevLett.96.180501}
}

@article{collins2000nmr,
  title={NMR quantum computation with indirectly coupled gates},
  author={Collins, David and Kim, Ki Wook and Holton, William C and Sierzputowska-Gracz, Hanna and Stejskal, EO},
  journal={Physical Review A},
  volume={62},
  number={2},
  pages={022304},
  year={2000},
  publisher={APS}
}

@article{wu2011experimental,
  title={Experimental demonstration of the Deutsch-Jozsa algorithm in homonuclear multispin systems},
  author={Wu, Zhen and Li, Jun and Zheng, Wenqiang and Luo, Jun and Feng, Mang and Peng, Xinhua},
  journal={Physical Review A},
  volume={84},
  number={4},
  pages={042312},
  year={2011},
  publisher={APS}
}

@article{takeuchi2000experimental,
  title={Experimental demonstration of a three-qubit quantum computation algorithm using a single photon and linear optics},
  author={Takeuchi, Shigeki},
  journal={Physical review A},
  volume={62},
  number={3},
  pages={032301},
  year={2000},
  publisher={APS}
}

@article{gulde2003implementation,
  title={Implementation of the Deutsch--Jozsa algorithm on an ion-trap quantum computer},
  author={Gulde, Stephan and Riebe, Mark and Lancaster, Gavin PT and Becher, Christoph and Eschner, J{\"u}rgen and H{\"a}ffner, Hartmut and Schmidt-Kaler, Ferdinand and Chuang, Isaac L and Blatt, Rainer},
  journal={Nature},
  volume={421},
  number={6918},
  pages={48--50},
  year={2003},
  publisher={Nature Publishing Group UK London}
}

@article{nagata2017quantum,
  title={Quantum cryptography based on the Deutsch-Jozsa algorithm},
  author={Nagata, Koji and Nakamura, Tadao and Farouk, Ahmed},
  journal={International Journal of Theoretical Physics},
  volume={56},
  pages={2887--2897},
  year={2017},
  publisher={Springer}
}

@article{biamonte2017quantum,
  title={Quantum machine learning},
  author={Biamonte, Jacob and Wittek, Peter and Pancotti, Nicola and Rebentrost, Patrick and Wiebe, Nathan and Lloyd, Seth},
  journal={Nature},
  volume={549},
  number={7671},
  pages={195--202},
  year={2017},
  publisher={Nature Publishing Group UK London}
}

@article{aaronson2021open,
  title={Open problems related to quantum query complexity},
  author={Aaronson, Scott},
  journal={ACM Transactions on Quantum Computing},
  volume={2},
  number={4},
  pages={1--9},
  year={2021},
  publisher={ACM New York, NY}
}

@article{nagata2015deutsch,
  title={The Deutsch-Jozsa algorithm can be used for quantum key distribution},
  author={Nagata, Koji and Nakamura, Tadao},
  journal={Open Access Library Journal},
  volume={2},
  number={8},
  pages={1--6},
  year={2015},
  publisher={Scientific Research Publishing}
}

@article{nagata2010can,
  title={Can von Neumann’s theory meet the Deutsch-Jozsa algorithm?},
  author={Nagata, Koji and Nakamura, Tadao},
  journal={International Journal of Theoretical Physics},
  volume={49},
  number={1},
  pages={162--170},
  year={2010},
  publisher={Springer}
}

@article{bishop2006pattern,
  title={Pattern recognition and machine learning},
  author={Bishop, Christopher M},
  journal={Springer google schola},
  volume={2},
  pages={645--678},
  year={2006}
}

@article{abbas2020quantum,
  title={On quantum ensembles of quantum classifiers},
  author={Abbas, Amira and Schuld, Maria and Petruccione, Francesco},
  journal={Quantum Machine Intelligence},
  volume={2},
  number={1},
  pages={6},
  year={2020},
  publisher={Springer}
}

@article{schuld2015introduction,
  title={An introduction to quantum machine learning},
  author={Schuld, Maria and Sinayskiy, Ilya and Petruccione, Francesco},
  journal={Contemporary Physics},
  volume={56},
  number={2},
  pages={172--185},
  year={2015},
  publisher={Taylor \& Francis}
}

@article{scarani2009security,
  title={The security of practical quantum key distribution},
  author={Scarani, Valerio and Bechmann-Pasquinucci, Helle and Cerf, Nicolas J and Du{\v{s}}ek, Miloslav and L{\"u}tkenhaus, Norbert and Peev, Momtchil},
  journal={Reviews of modern physics},
  volume={81},
  number={3},
  pages={1301--1350},
  year={2009},
  publisher={APS}
}

@book{wittek2014quantum,
  title={Quantum machine learning: what quantum computing means to data mining},
  author={Wittek, Peter},
  year={2014},
  publisher={Academic Press}
}

@article{nagata2020some,
  title={Some theoretically organized algorithm for quantum computers},
  author={Nagata, Koji and Nakamura, Tadao},
  journal={International Journal of Theoretical Physics},
  volume={59},
  number={2},
  pages={611--621},
  year={2020},
  publisher={Springer}
}

@article{nagata2020generalization,
  title={Generalization of Deutsch’s algorithm},
  author={Nagata, Koji and Nakamura, Tadao},
  journal={International Journal of Theoretical Physics},
  volume={59},
  pages={2557--2561},
  year={2020},
  publisher={Springer}
}

@article{Preskill2018,
  title={Quantum Computing in the NISQ era and beyond},
  author={Preskill, John},
  journal={Quantum},
  volume={2},
  pages={79},
  year={2018},
  url={https://quantum-journal.org/papers/v2/q-2018-08-06-79/}
}

@book{NielsenChuang,
  title={Quantum Computation and Quantum Information},
  author={Nielsen, Michael A. and Chuang, Isaac L.},
  year={2010},
  publisher={Cambridge University Press},
  edition={10th Anniversary Edition}
}

@article{Bharti2022,
  title={Noisy intermediate-scale quantum (NISQ) algorithms},
  author={Bharti, Kishor and Cervera-Lierta, Alvaro and Kyaw, Thi Ha and Haug, Tobias and Alsing, Paul and Aspuru-Guzik, Alan},
  journal={Reviews of Modern Physics},
  volume={94},
  issue={1},
  pages={015004},
  year={2022},
  url={https://journals.aps.org/rmp/abstract/10.1103/RevModPhys.94.015004}
}

@article{Berry1941,
  title = {The accuracy of the Gaussian approximation to the sum of independent variates},
  author = {Berry, A. C.},
  journal = {Transactions of the American Mathematical Society},
  volume = {49},
  number = {1},
  pages = {122--136},
  year = {1941},
  publisher = {American Mathematical Society},
  doi = {10.2307/1990055},
  url = {https://doi.org/10.2307/1990055}
}

@article{Esseen1942,
  title = {On the Liapunoff limit of error in the theory of probability},
  author = {Esseen, C.-G.},
  journal = {Arkiv f\"or Matematik, Astronomi och Fysik},
  volume = {28A},
  number = {9},
  pages = {1--19},
  year = {1942},
  publisher = {Kungl. Svenska Vetenskapsakademien}
}

@book{Petrov1995,
  title = {Limit Theorems of Probability Theory: Sequences of Independent Random Variables},
  author = {Petrov, V. V.},
  year = {1995},
  publisher = {Oxford University Press},
  address = {Oxford, UK},
  isbn = {978-0-19-853499-4}
}

@inproceedings{bennett1984bb84,
  title        = {Quantum cryptography: Public key distribution and coin tossing},
  author       = {Bennett, Charles H. and Brassard, Gilles},
  booktitle    = {Proceedings of IEEE International Conference on Computers, Systems and Signal Processing},
  pages        = {175--179},
  year         = {1984},
  address      = {Bangalore, India},
  publisher    = {IEEE},
  url          = {https://doi.org/10.48550/arXiv.2003.06557}
}

@book{CoverThomas2006,
  author    = {Thomas M. Cover and Joy A. Thomas},
  title     = {Elements of Information Theory},
  edition   = {2nd},
  publisher = {Wiley},
  year      = {2006},
  note      = {See chs. on hypothesis testing, Chernoff information, and large deviations}
}

@article{FuchsPeres1996,
  author  = {Christopher A. Fuchs and Asher Peres},
  title   = {Quantum-state disturbance versus information gain: Uncertainty relations for quantum information},
  journal = {Phys. Rev. A},
  volume  = {53},
  number  = {4},
  pages   = {2038--2045},
  year    = {1996},
  doi     = {10.1103/PhysRevA.53.2038}
}

@article{Scarani2009RMP,
  author  = {Valerio Scarani and Helle Bechmann-Pasquinucci and Nicolas J. Cerf and Miloslav Du\v{s}ek and Norbert L\"utkenhaus and Momtchil Peev},
  title   = {The security of practical quantum key distribution},
  journal = {Rev. Mod. Phys.},
  volume  = {81},
  pages   = {1301--1350},
  year    = {2009},
  doi     = {10.1103/RevModPhys.81.1301}
}

@article{Kawakami2017Bernoulli,
  author    = {Shun Kawakami and Toshihiko Sasaki and Masato Koashi},
  title     = {Finite-key analysis for quantum key distribution with weak coherent pulses based on Bernoulli sampling},
  journal   = {Physical Review A},
  volume    = {96},
  number    = {1},
  pages     = {012305},
  year      = {2017},
  doi       = {10.1103/PhysRevA.96.012305},
  url       = {https://doi.org/10.1103/PhysRevA.96.012305},
  eprint    = {1701.04168},
  archivePrefix = {arXiv},
  primaryClass  = {quant-ph}
}

@article{Maeda2019FiniteKey,
  author    = {Kento Maeda and Toshihiko Sasaki and Masato Koashi},
  title     = {Repeaterless quantum key distribution with efficient finite-key analysis overcoming the rate-distance limit},
  journal   = {Nature Communications},
  year      = {2019},
  volume    = {10},
  number    = {1},
  pages     = {3140},
  doi       = {10.1038/s41467-019-11008-z},
  url       = {https://www.nature.com/articles/s41467-019-11008-z}
}

@misc{Koashi2020Tutorial,
  author       = {Masato Koashi},
  title        = {Tutorial: Security of quantum key distribution: approach from complementarity},
  howpublished = {QCrypt 2020 tutorial slides},
  institution  = {University of Tokyo},
  year         = {2020},
  month        = {August},
  url          = {https://2020.qcrypt.net/slides/QCrypt2020_Koashi_Tutorial.pdf},
  note         = {Slides discuss Bernoulli sampling and finite-key analysis in QKD}
}

@article{chen2020characterization,
  title={Characterization of exact one-query quantum algorithms},
  author={Chen, Weijiang and Ye, Zekun and Li, Lvzhou},
  journal={Physical Review A},
  volume={101},
  number={2},
  pages={022325},
  year={2020},
  publisher={APS}
}

@article{bennett1992b92,
  title        = {Quantum cryptography using any two nonorthogonal states},
  author       = {Bennett, Charles H.},
  journal      = {Physical Review Letters},
  volume       = {68},
  number       = {21},
  pages        = {3121--3124},
  year         = {1992},
  doi          = {10.1103/PhysRevLett.68.3121}
}

@article{ekert1991e91,
  title        = {Quantum cryptography based on Bell’s theorem},
  author       = {Ekert, Artur K.},
  journal      = {Physical Review Letters},
  volume       = {67},
  number       = {6},
  pages        = {661--663},
  year         = {1991},
  doi          = {10.1103/PhysRevLett.67.661}
}

@article{wootters1982single,
  title        = {A single quantum cannot be cloned},
  author       = {W.~K. Wootters and W.~H. Zurek},
  journal      = {Nature},
  volume       = {299},
  number       = {5886},
  pages        = {802--803},
  year         = {1982},
  doi          = {10.1038/299802a0}
}

\clearpage
\appendix
%%%%%%%%%%
%\subsection{Generalized D-J algorithm calculations}

\section*{Cloud platform details}
In accordance with the Physical Review policy on cloud quantum computing demonstrations,\footnote{\url{https://journals.aps.org/pra/edannounce/cloud-quantum-computing-demonstrations-pra}} we document the platform as follows.
%\begin{itemize}
    \paragraph{\textbf{Provider / backend}: IBM Quantum, \texttt{ibm-brisbane} (127-qubit Falcon family backend).}
    \paragraph{\textbf{Access mode}: Public cloud quantum hardware accessed via Qiskit.}
    \paragraph{\textbf{Software stack}: Qiskit (version recorded in code repository), Python 3.x. Source code is available at \url{https://github.com/MiladGhadimi/Generalized-Deutsch-Josza-Quantum-Algorithm}.}
    \paragraph{\textbf{Qubit resources actually used}: circuits used four data qubits plus one ancilla for GD/GDJ demonstrations (see Figs. 8-10).}
    \paragraph{\textbf{Shots}: Unless otherwise noted, each circuit execution used 4000 shots.}
    \paragraph{\textbf{Dates of access}: Demonstrations performed on publicly available backends prior to submission; backend names and code commit hashes are recorded in the repository to ensure reproducibility.}
    \paragraph{\textbf{Noise / calibration notes}: As standard for cloud hardware, gate and readout errors, coherence times, and qubit connectivity vary over time; the calibration snapshot at run time is stored in the job metadata (retrieved via Qiskit) and referenced in the repository logs.}
%\end{itemize}

%All results in this work are therefore described as \emph{cloud-based demonstrations} on public quantum hardware rather than laboratory experiments. 

\subsubsection{Algorithm simulation and verification using IBM quantum computer}

IBM Quantum provides access to some of the most advanced quantum computers available today, along with a comprehensive suite of tools and libraries through the Qiskit framework. Qiskit, an open-source quantum computing software development framework to perform quantum algorithms and run simulations on classical computers as well as execute them on real quantum hardware as mentioned aboe.
%In this study, we used an IBM Quantum computer, specifically the $\text{ibm}_{-}\text{brisbane}$ quantum processor, featuring a 127-qubit system. 
Our aim was to implement the Generalized Deutsch (GD) and Generalized Deutsch-Jozsa (GDJ) algorithms, employing a configuration of four qubits for input and an ancillary qubit to construct the oracle, as depicted in the circuit diagram in FIG.\ref{fig4q}, but her frist we start with two qubit or GD algorithm them develop it for GDJ algorithm as the results represented in Fig.\ref{res4q}. For detalits of the codes it exists in GitHub: \url{https://github.com/MiladGhadimi/Generalized-Deutsch-Josza-Quantum-Algorithm}

%\subsection{Implementation of GD algorithm on Qiskit}

Here, the results of demonstrating our algorithm on Qiskit software have been demonstrated. 
\begin{figure*}[htbp]
\centering
\includegraphics[width=0.46\textwidth]{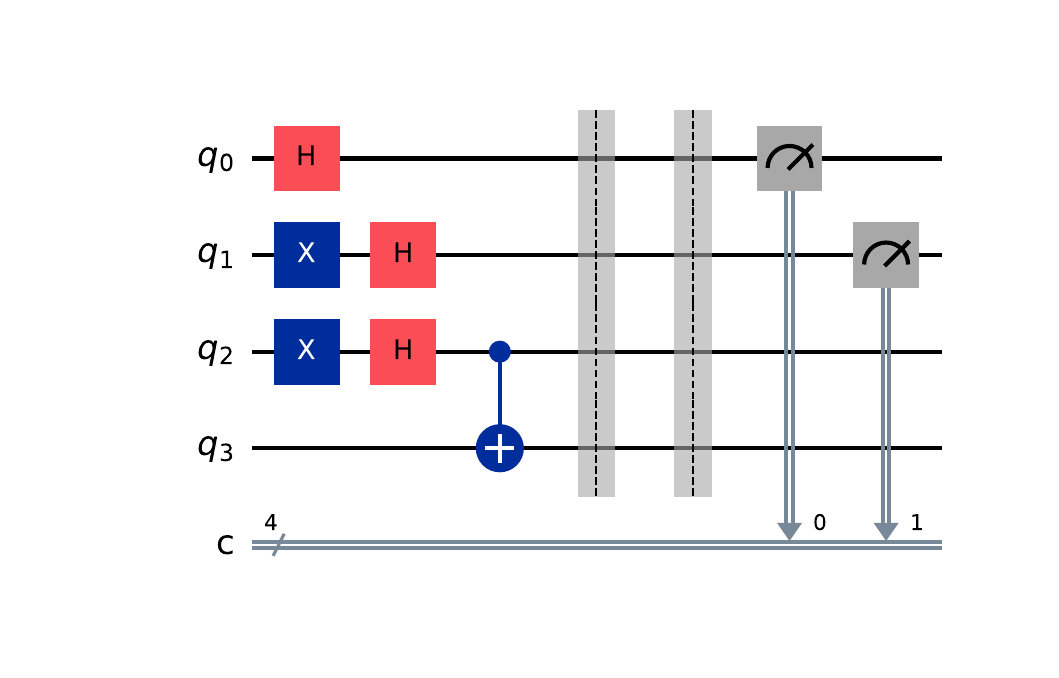}%
\hfill
\includegraphics[width=0.5\textwidth]{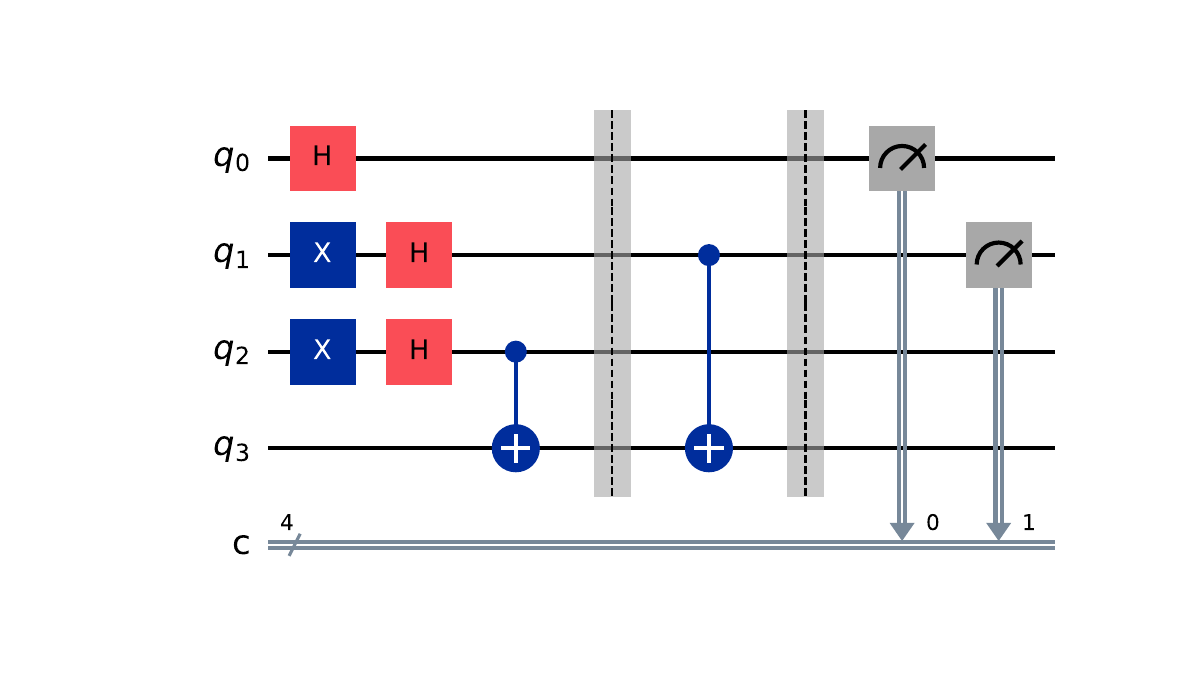}
\hfill
\includegraphics[width=0.35\textwidth]{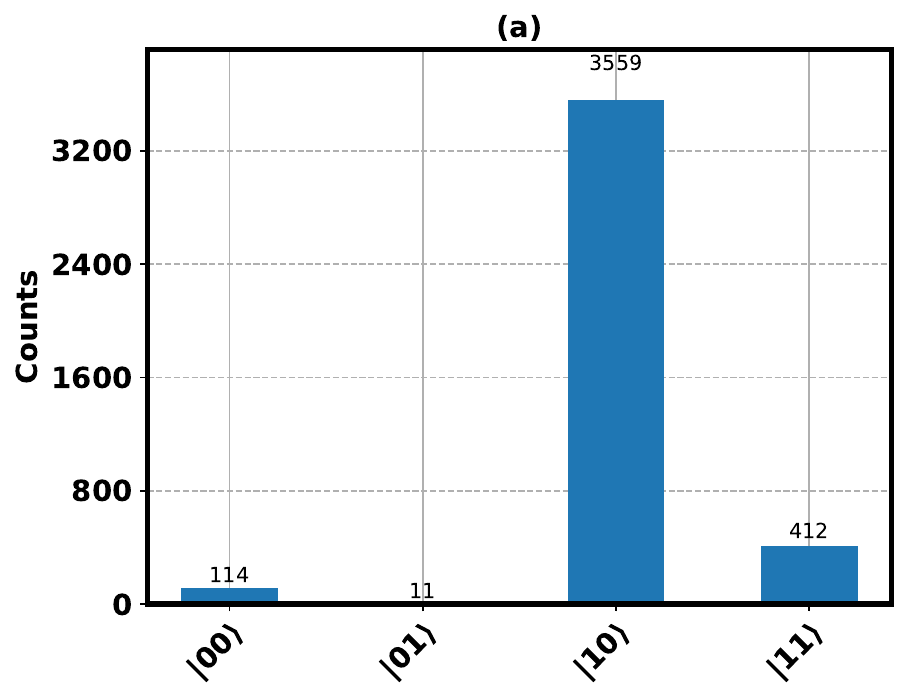}%
\hfill
\includegraphics[width=0.35\textwidth]{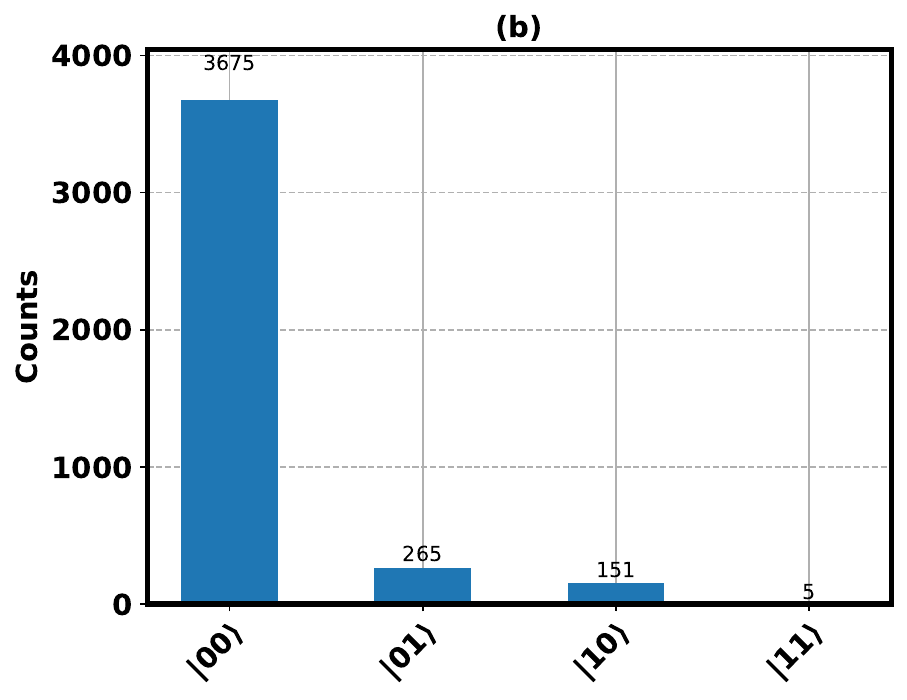}
\hfill
\includegraphics[width=0.46\textwidth]{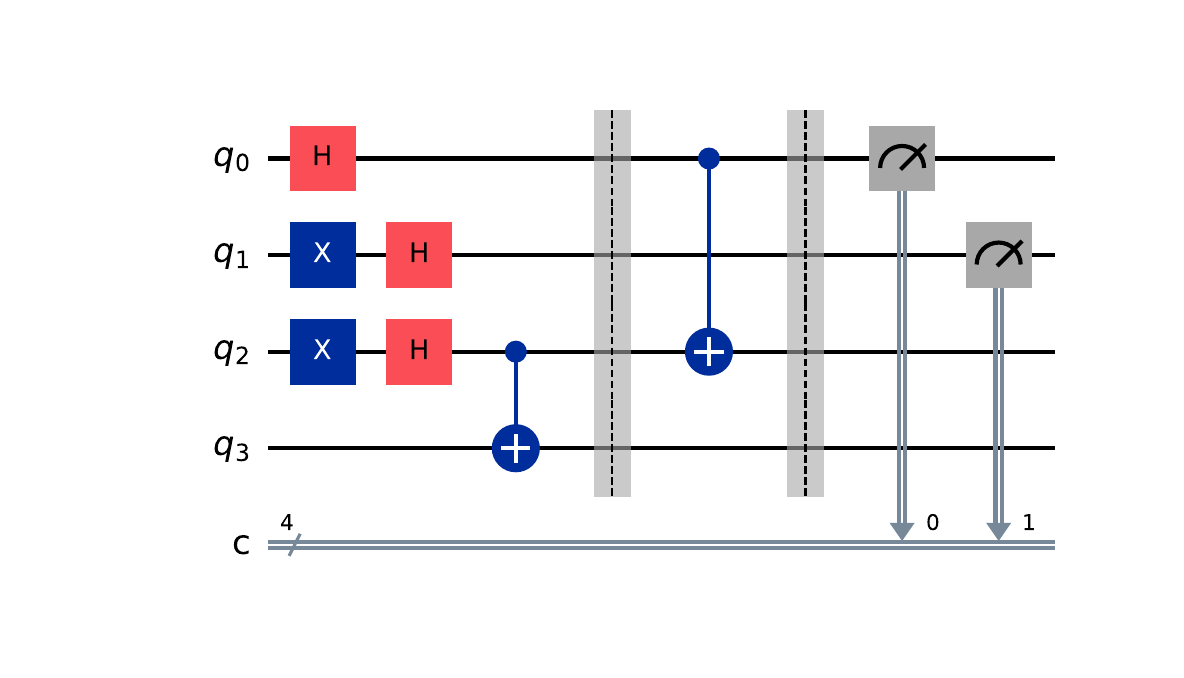}%
\hfill
\includegraphics[width=0.5\textwidth]{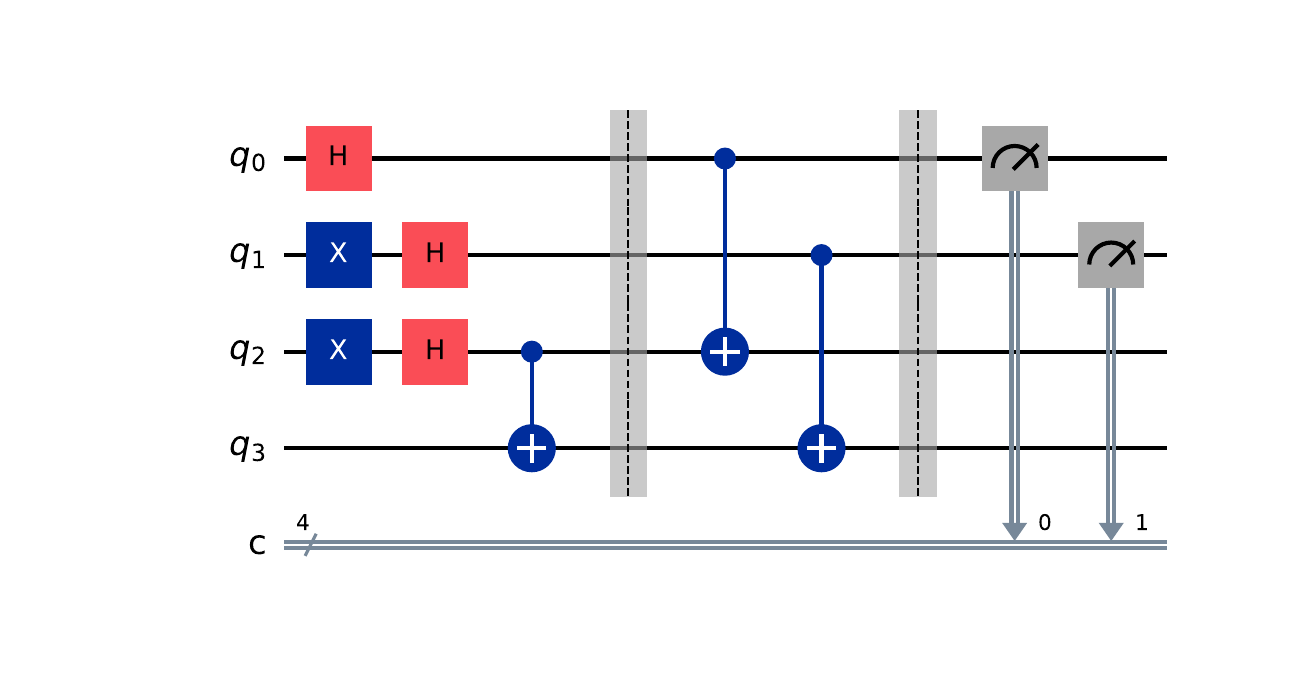}
\hfill
\includegraphics[width=0.35\textwidth]{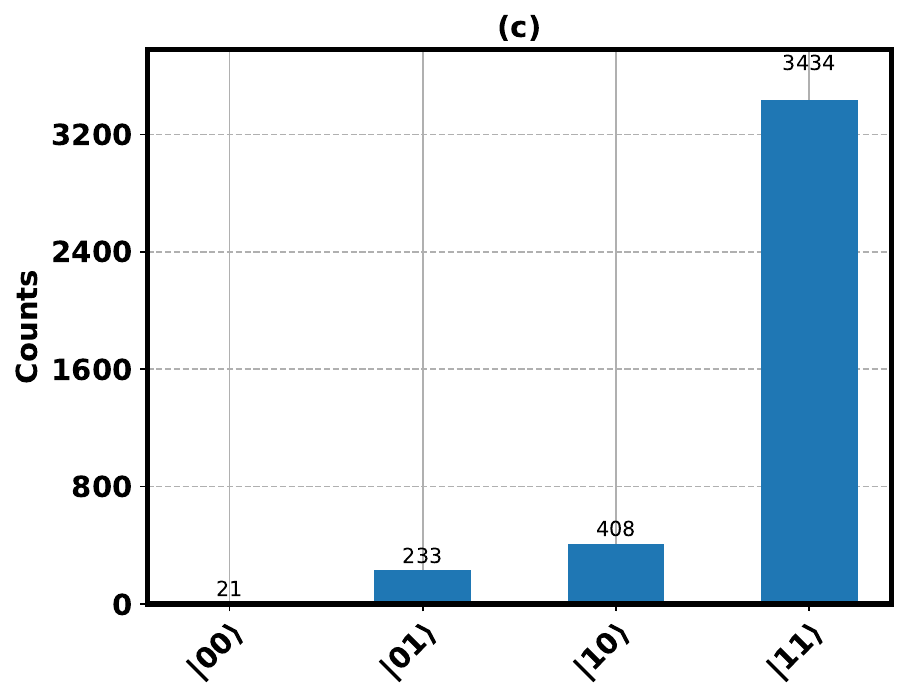}%
\hfill
\includegraphics[width=0.35\textwidth]{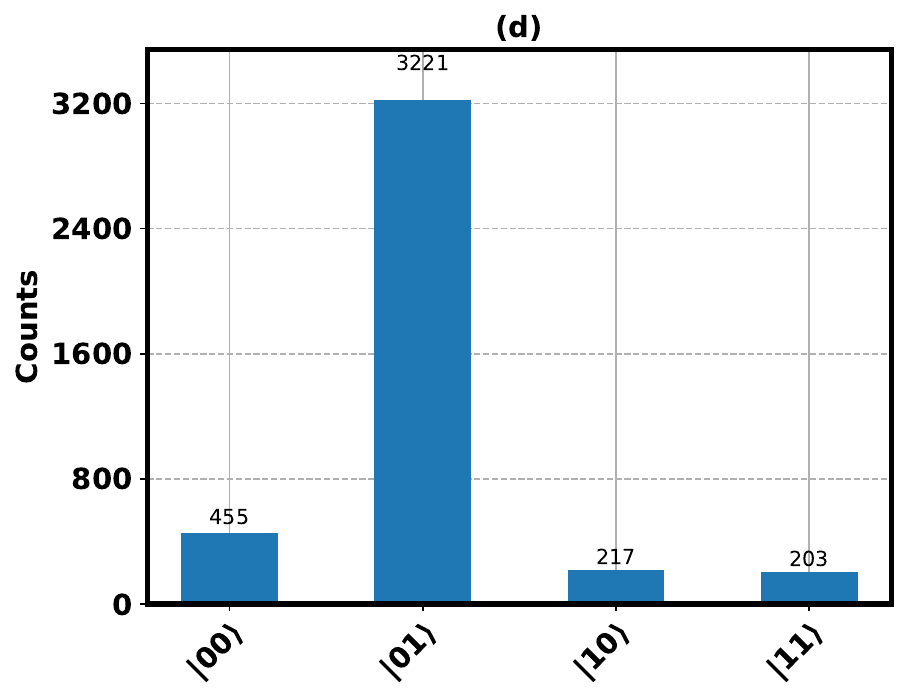}
\caption{Illustration of all possible functions for the Generalized Deutsch Algorithm: Top row, left to right, Circiut for constant functions,   balanced functions; second row, left to right, (a) Function is constant $f(0)=0, f(1)=0$ (basis results), (b) Function is balanced $f(0)=0, f(1)=1$ (basis results); third row, left to right, circiut for balanced functions $f(0)=1, f(1)=0$,  and constant $f(0)=f(1)=1$; bottom row, left to right, (c) Function is balanced $f(0)=1, f(1)=0$ (basis results), (d) Function is constant $f(0)=f(1)=1$ (basis results).}\label{fig9}
\end{figure*}

% \iffalse

% \begin{figure*}[h!]
% \centering
% \includegraphics[width=0.7\textwidth]{c.pdf}
% \caption{IBM Quantum Circuit for GDJ algorithm based on four qubits}\label{figg}
% \end{figure*}

% \begin{figure*}[htbp]
% \centering
% \includegraphics[width=0.35\textwidth]{i11.png}%
% \hfill % Use \hfill to separate the images as needed
% \includegraphics[width=0.35\textwidth]{i22.png}\\ % Creates a new line for the next pair of figures
% \includegraphics[width=0.35\textwidth]{i33.png}%
% \hfill % Use \hfill to separate the images as needed
% \includegraphics[width=0.35\textwidth]{i44.png}
% \caption{(a) Function is balanced $f(0)=0, f(1)=1$. (b) Function is balanced $f(0)=1, f(1)=0$. (c) Function is constant $f(0)=f(1)=1$. (d) Function is constant $f(0)=f(1)=0$.}\label{fig9}
% \end{figure*}

% \fi

\begin{figure*}[h!]
\centering
\includegraphics[width=0.85\textwidth]{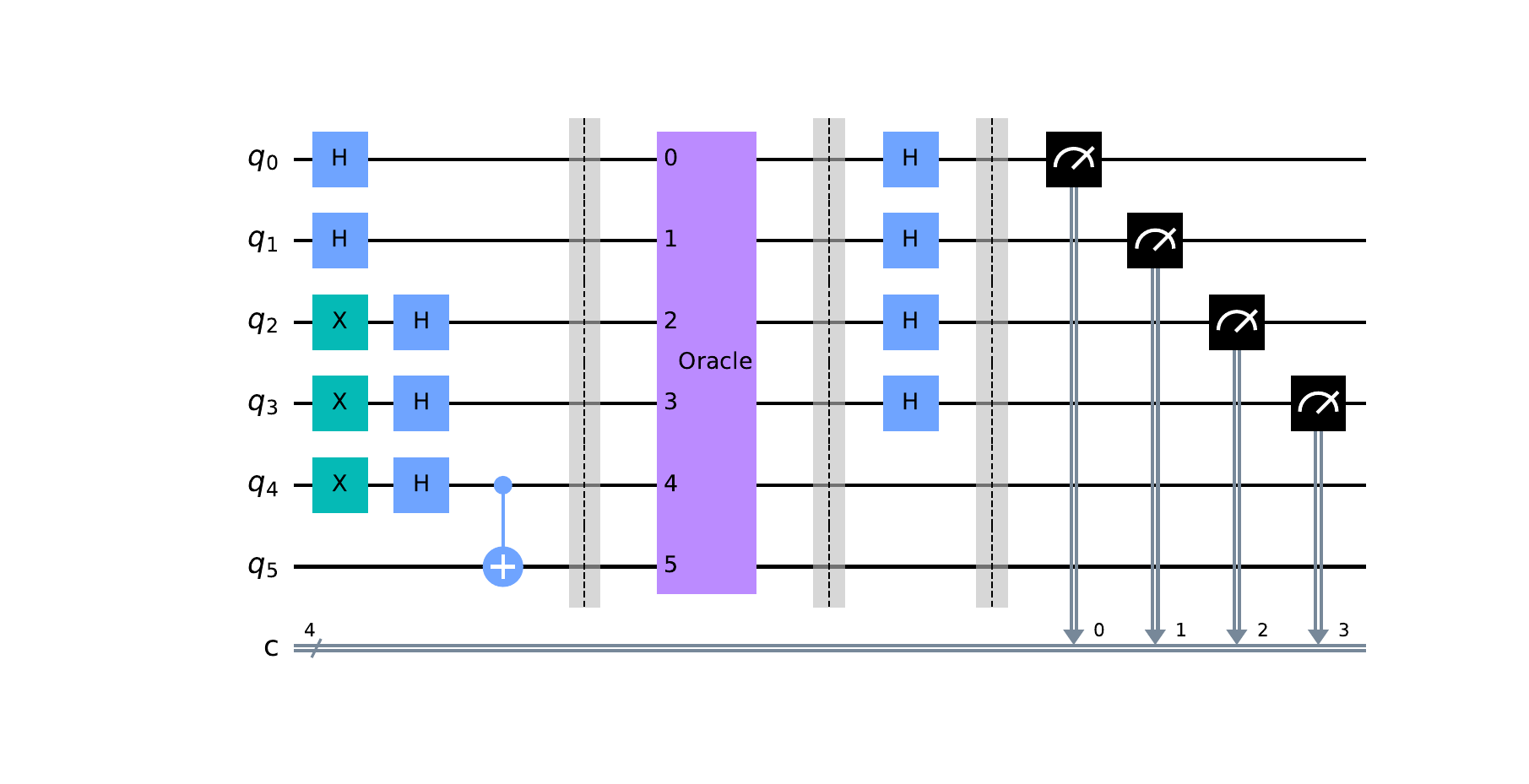}
\caption{IBM Quantum Circuit for GDJ algorithm based on four qubits.}\label{fig4q}
\end{figure*}

\begin{figure*}[htbp]
\centering
%\vspace{-0.6cm}
\includegraphics[width=0.35\textwidth]{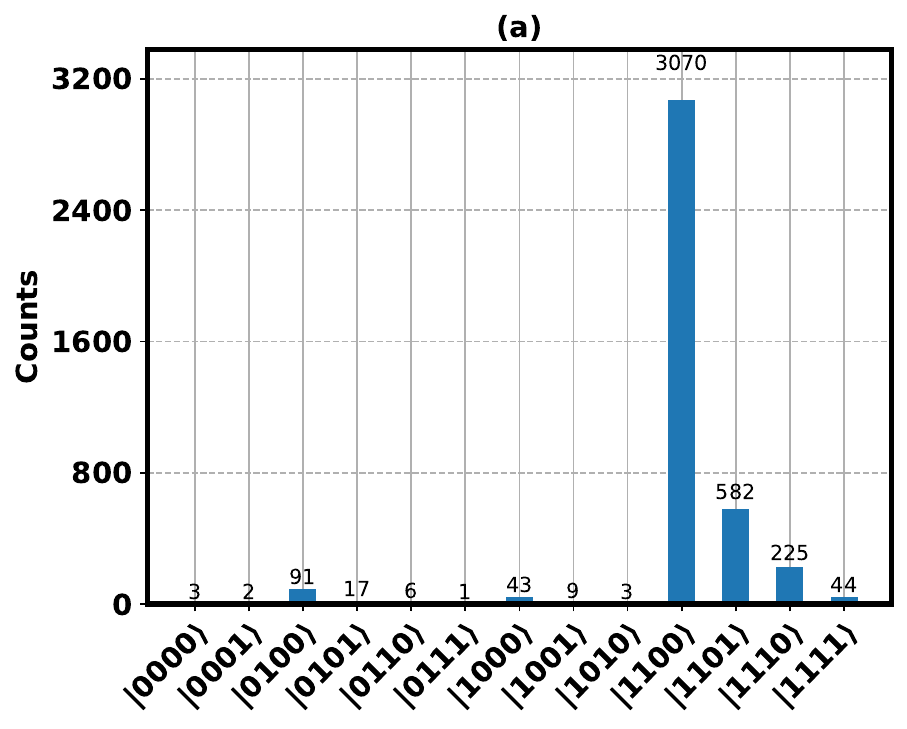}%
\hfill
%\vspace{0.9cm}% Use \hfill to separate the images as needed
\includegraphics[width=0.35\textwidth]{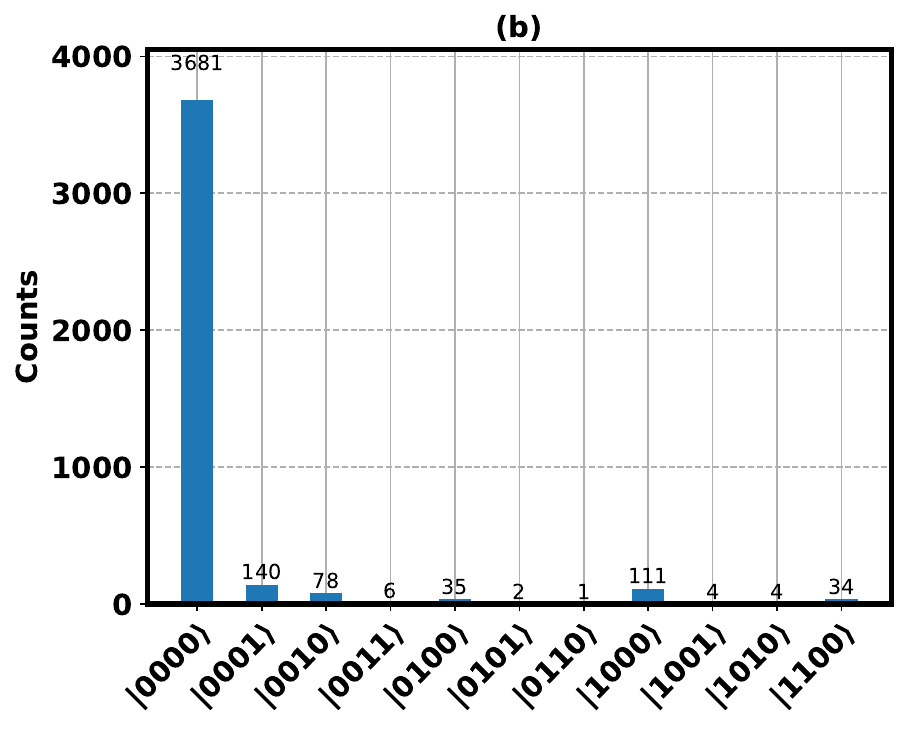}
\hfill
%\vspace{0.1cm}
% Creates a new line for the next pair of figures
\includegraphics[width=0.35\textwidth]{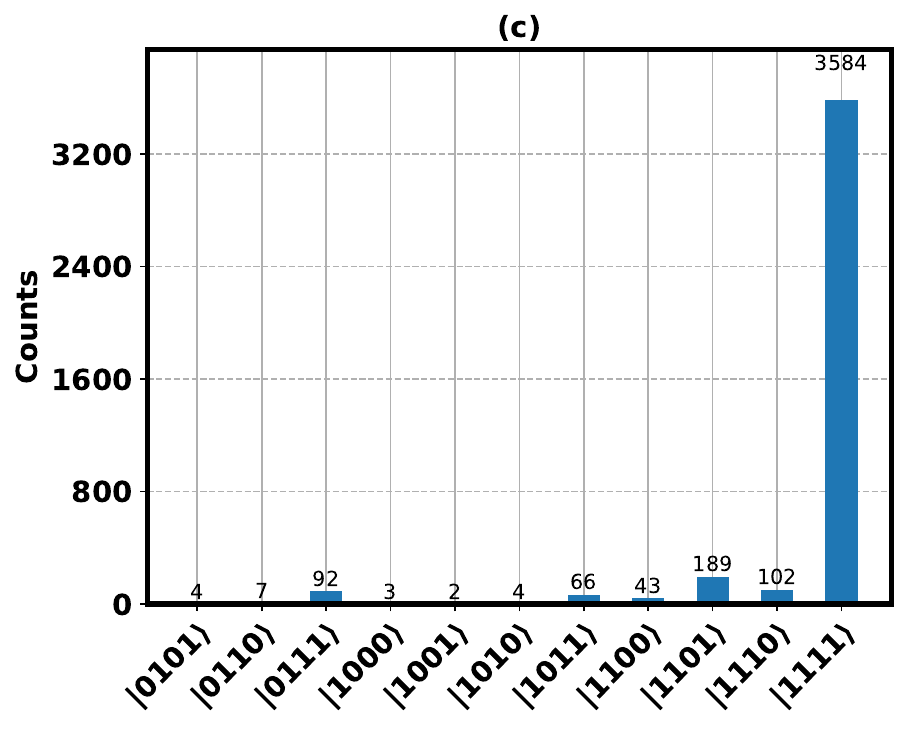}%
\hfill % Use \hfill to separate the images as needed
\includegraphics[width=0.35\textwidth]{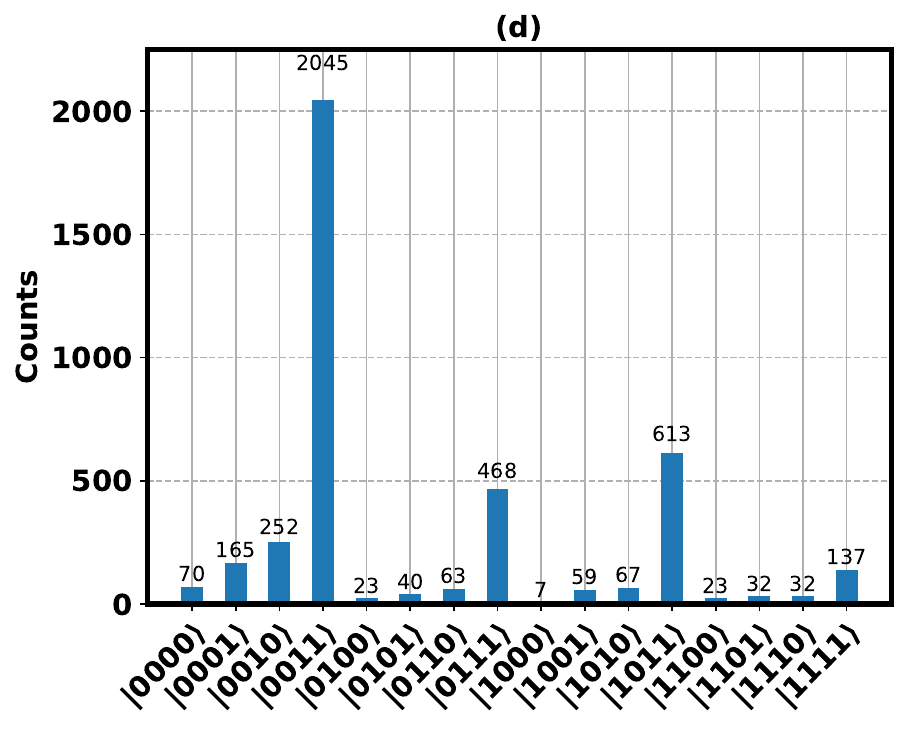}
\vspace{0.5cm}
\caption{(a) Function is constant $f(0)=f(1)=0$. (b)Function is constant $f(0)=f(1)=1$ (c) Function is balanced $f(0)=1, f(1)=0$. (d) Function is balanced $f(0)=0, f(1)=1$.}\label{res4q}

\end{figure*}

\section{Implementation of GD and GDJ algorithm on Qiskit and IBMQ}
Executing the algorithm on the IBM Quantum platform yielded results showcased in FIG.\ref{fig9}, illustrating distinctive outcomes for each type of oracle—\textbf{constant 0}, \textbf{constant 1}, \textbf{balanced 01}, and \textbf{balanced 10}. These results were obtained through a total of 4000 shots on the quantum processor, aiming for statistically significant results to verify theoretical predictions and simulations. Our quantum circuit was precisely designed to specify the exact value of the function, highlighting the effectiveness of our implementation on a quantum computing platform in differentiating constant and balanced functions.
We begin by initializing our quantum circuit. In this phase, we set up the required qubits and classical bits. For the Generalized Deutsch (GD) algorithm, we configure two qubits for input and an ancillary qubit $|\Phi^-\rangle$ to construct the oracle. Additionally, we allocate classical bits to store measurement results.

%\subsection{Step 2: Creating the Oracle}
The oracle is a crucial component of the GD algorithm. It encodes the function we're analyzing into the quantum state. In our implementation, we'll design the oracle to represent a specific function. Depending on the function's nature—whether it's constant or balanced—the oracle's structure will vary. For general case when both are constant and $0$ we do not put anything as oracle, but for any change for balancing and changing the function to $1$ we only use CNOT gate between the each input and $|\Phi^-\rangle$ ancillary.

%\subsection{Step 3: Implementing Quantum Gates}
In this phase, we apply quantum gates to manipulate the quantum state of the qubits. These gates are the building blocks of quantum algorithms, enabling operations such as superposition, entanglement, and interference. For the GDJ algorithm, we'll use gates like Hadamard gates.

%\subsection{Step 4: Measurement}
After applying the necessary quantum operations, we perform measurements on \text{q0 and q1} to extract information from the quantum system. The measurement results are stored in classical bits, allowing us to analyze the outcome of the algorithm FIG.\ref{fig9}. By repeating the measurement process multiple $1024$ times, we gather statistical data to validate the algorithm's behavior and verify its correctness.

%\subsection{Step 5: Result Analysis}
Once we've obtained measurement results FIG.\ref{fig9}, we analyze them to draw conclusions about the function being evaluated. As we simulated the function for all possible version as constant with its value or balanced with basis results. The distinctive patterns observed in the measurement results provide insights into the algorithm's ability to differentiate different types of functions. The running of GDJ algorithm with 4 qubits are demonstrated in Figs. 9-\ref{res4q}.

\section{Quantum Cryptography: Chernoff bound, Chernoff information and rate constant}

\noindent\textbf{Chernoff bound.} A moment-generating-function tail bound for sums of independent random variables. Let $X_1,\dots,X_n$ be independent real-valued random variables and let $S_n=\sum_{i=1}^n X_i$ is their sum over $n\in\mathbb{N}$ trials. For any threshold level $t\in\mathbb{R}$,
\[
\Pr(S_n\ge n t)\;\le\;e^{-n\Lambda^{*}(t)},
\]
where $\Lambda^{*}(t)$ is the convex conjugate (rate function) of the log-moment-generating function $\Lambda(\lambda)=\log\mathbb{E}[e^{\lambda X_1}]$ (i.e., $\Lambda^{*}(t)=\sup_{\lambda>0}\{\lambda t-\Lambda(\lambda)\}$). This tail bound implies the exponential miss-probability decay used in Eq.~(31) for threshold tests on the flag count $S$ \cite{CoverThomas2006}.

\noindent\textbf{Chernoff information (rate) $C(P,Q)$.} In binary hypothesis testing between two distributions $P$ and $Q$ over an alphabet $\mathcal{X}$ (with generic element $x\in\mathcal{X}$), the optimal Bayes error after $n$ i.i.d.\ samples satisfies
\[
P_{\mathrm{err}}^{*}(n)\;\asymp\;e^{-n\,C(P,Q)},
\]

%\qquad
\[
C(P,Q)\;=\;-\min_{0\le s\le 1}\,\log\!\sum_{x\in\mathcal{X}} P(x)^{s}Q(x)^{1-s},
\]
where $\asymp$ means asymptotic equivalence in exponential rate (same leading decay/growth exponent), and $s\in[0,1]$ is the Chernoff exponent. For Bernoulli$(q_0)$ vs.\ Bernoulli$(q_1)$ (one-bit outcomes with success probabilities $q_0,q_1\in(0,1)$), this yields $k:=C(q_0,q_1)$ as the constant in the detection law $P_{\mathrm{detect}}(d)\approx 1-e^{-k\,\eta d}$ \cite{CoverThomas2006}.\\

\noindent\textbf{Rate constant $k$.} The exponent in Eq.~(31) controlling how fast detection improves with blocklength. In the hypothesis-testing view, $k=C(q_0,q_1)$, the Chernoff information between the honest per-trial flag law $\mathrm{Bern}(q_0)$ and the attacked law $\mathrm{Bern}(q_1)$. In the independent-trials view, if a single attacked trial is caught with probability $\alpha\in(0,1)$, then $k=-\ln(1-\alpha)\approx\alpha$ for small $\alpha$. This packages device parameters and attack strength into a single slope governing detection vs.\ $d$.

\medskip
\emph{Detection scaling.} Let $d=2^{n}$ be the number of transmissions (effective dimension) and let $\eta\in[0,1]$ be the attacked fraction, so $m=\eta d$ trials are attacked. Large-deviation/Chernoff theory for sums of Bernoulli flags gives
\[
\Pr[\text{miss Eve}] \;\lesssim\; \exp\!\big(-\,C(q_0,q_1)\,m\big),
\]
hence the detection probability
\[
P_{\mathrm{detect}}(d)=1-\Pr[\text{miss Eve}] \;\gtrsim\; 1-\exp\!\big(-\,C(q_0,q_1)\,\eta d\big),
\]
which is Eq.~(31) with $k:=C(q_0,q_1)$, where $q_0$ is the per-trial flag probability with no eavesdropper (honest channel) and $q_1>q_0$ is the per-trial flag probability on an attacked trial \cite{CoverThomas2006}. Equivalently, if an attacked trial is caught with probability $\alpha$, independence across the $m$ attacked trials gives
\[
P_{\mathrm{detect}}(d)=1-(1-\alpha)^{\eta d}\;\approx\; 1-e^{-\alpha\,\eta d},
\]
which matches Eq.~(31) for $k=\alpha$ (Poisson approximation valid for small $\alpha$).

\medskip
\emph{GDJ amplifies disturbance:} Under intercept–resend, Eve measures and re-sends, erasing the relative phases created by the oracle and Bell ancilla; after Bob’s Hadamards, outcomes are nearly uniform over the admissible symbols \cite{Scarani2009RMP}. For DJ there are $2$ symbols, so the accidental match (no flag) probability is $1/2$, giving the per-trial catch $\alpha_{\mathrm{DJ}}=\tfrac{1}{2}$. For GDJ there are $4$ symbols, so the accidental match is $1/4$, giving $\alpha_{\mathrm{GDJ}}=\tfrac{3}{4}$. Therefore
\[
P_{\mathrm{detect}}^{\mathrm{DJ}}(d)\approx 1-e^{-(1/2)\,\eta d},\qquad
P_{\mathrm{detect}}^{\mathrm{GDJ}}(d)\approx 1-e^{-(3/4)\,\eta d},
\]
so GDJ boosts the exponent by a factor $\approx 3/2$ in this idealized setting. Real devices include background error $q_0>0$, but the information–disturbance tradeoff still ensures $q_1>q_0$ (and thus $k>0$) \cite{FuchsPeres1996,Scarani2009RMP}.

\end{document}